\documentclass[a4paper,fleqn]{cas-dc}

\usepackage[numbers]{natbib}

\usepackage{algorithm}
\usepackage{algorithmic}
\usepackage{caption}

\usepackage[caption=false,font=normalsize,labelfont=sf,textfont=sf]{subfig}

\def\tsc#1{\csdef{#1}{\textsc{\lowercase{#1}}\xspace}}
\tsc{WGM}
\tsc{QE}
\tsc{EP}
\tsc{PMS}
\tsc{BEC}
\tsc{DE}

\begin{document}
\let\WriteBookmarks\relax
\def\floatpagepagefraction{1}
\def\textpagefraction{.001}
\shorttitle{Preprint}
\shortauthors{J. Duan et~al.}

\title [mode = title]{Adaptive dynamic programming for nonaffine nonlinear optimal control problem with state constraints}

\author[1]{Jingliang Duan}
\ead{duanjl15@163.com}

\address[1]{State Key Lab of Automotive Safety and Energy, School of Vehicle and Mobility, Tsinghua University, Beijing, 100084, China.}

\author[1]{Zhengyu Liu}
\ead{liuzheng17@mails.tsinghua.edu.cn}

\author[1]{Shengbo Eben Li}[orcid=0000-0003-4923-3633]
\cormark[1]
\ead{lishbo@tsinghua.edu.cn}

\author[1]{Qi Sun}
\ead{qisun@tsinghua.edu.cn}

\author[2]{Zhenzhong Jia}
\ead{zhenzhong.jia@gmail.com}

\address[2]{Robotics Institute at Carnegie Mellon University, Pittsburgh, PA 15213, USA. }

\author[1]{Bo Cheng}
\ead{chengbo@tsinghua.edu.cn}

\cortext[cor1]{Corresponding author}

\begin{abstract}
This paper presents a constrained adaptive dynamic programming (CADP) algorithm to solve general nonlinear nonaffine optimal control problems with known dynamics. Unlike previous ADP algorithms, it can directly deal with problems with state constraints. Firstly, a constrained generalized policy iteration (CGPI) framework is developed to handle state constraints by transforming the traditional policy improvement process into a constrained policy optimization problem. Next, we propose an actor-critic variant of CGPI, called CADP, in which both policy and value functions are approximated by multi-layer neural networks to directly map the system states to control inputs and value function, respectively. CADP linearizes the constrained optimization problem locally into a quadratically constrained linear programming problem, and then obtains the optimal update of the policy network by solving its dual problem. A trust region constraint is added to prevent excessive policy update, thus ensuring linearization accuracy. We determine the feasibility of the policy optimization problem by calculating the minimum trust region boundary and update the policy using two recovery rules when infeasible. The vehicle control problem in the path-tracking task is used to demonstrate the effectiveness of this proposed method. 
\end{abstract}


\maketitle

\newdefinition{definition}{Definition}
\newtheorem{lemma}{Lemma}
\newtheorem{theorem}{Theorem}
\newtheorem{assumption}{Assumption}
\newtheorem{remark}{Remark}
\newproof{proof}{Proof}

\section{Introduction}
Dynamic programming (DP) is a theoretical and effective tool in solving discrete-time (DT) optimal control problems with known dynamics \cite{bertsekas2005DP&OC}.  The optimal value function (or cost-to-go) for DT systems is obtained by solving the DT Hamilton-Jacobi-Bellman (HJB) equation, also known as the Bellman optimality equation, which develops backward in time \cite{lewis2012OptimalControl}. 
However, due to the curse of dimensionality,
running DP directly to get the optimal solution of DT HJB is usually computationally untenable for complex nonlinear DT systems \cite{wang2009adaptive}. The adaptive dynamic programming (ADP) methods were first proposed by Werbos as a way to overcome this difficulty by solving an approximate solution of DT HJB forward in time \cite{werbos1974phD,werbos1992approximate}. In some studies, ADP is also called approximate dynamic programming \cite{powell2007approximate,liu2017ADP}.

ADP methods are usually implemented as an actor-critic architecture which involves a critic parameterized function for value function approximation and an actor parameterized function for policy approximation \cite{duan2019generalized,liu2017ADP,vamvoudakis2010online,li2017off}. Neural networks (NNs) have been widely used as approximators of both value function and policy due to their strong fitting ability, and have achieved good performance on many control tasks \cite{liu2017ADP}. 
Most ADP methods adopt generalized policy iteration (GPI) as a primary tool to adjust both value and policy networks by iteratively solving the DT HJB equation \cite{liu2017ADP}. The well-known policy iteration and value iteration methods can also be regarded as special cases of GPI \cite{liu2017ADP,sutton2018reinforcement}. There are two revolving iteration procedures for GPI framework: 1) policy evaluation, which drives the value function towards the true value function for the current policy, and 2) policy improvement, which improves the policy to reduce the current value function.

Over the last few decades, many ADP methods of finding the nearly optimal control solution for DT systems with known dynamics have emerged. Chen and Jagannathan proposed an ADP  method to find nearly optimal control state feedback laws for input-affine nonlinear DT systems by iteratively solving the generalized HJB equation. The value function was approximated by a linear combination of artificially designed basis functions, while the policy was directly derived from the value function \cite{chen2008generalizedHJB}. Both actor and critic NNs were utilized by Al-Tamimi \emph{et al}. to develop a value-iteration-based algorithm for DT systems, and it was shown that the algorithm could converge to the optimal value function and policy as long as the control coefficient matrix was known \cite{al2008discrete}. Liu \emph{et al}. proposed a GPI algorithm for DT nonlinear systems, and the admissibility property of the policy network during learning could be guaranteed as long as the initialized policy was admissible \cite{liu2015generalized}.

ADP is usually considered together with reinforcement learning (RL) because both provide approximate solutions to DP \cite{sutton2018reinforcement}. Compared with ADP, RL focuses more on solving the nearly optimal policy of high-dimensional systems with completely unknown models, such as Atari games \cite{mnih2015human} and StarCraft \cite{vinyals2019grandmaster}.  In recent years, RL algorithms such as DSAC (Distributional Soft Actor-Critic) \cite{duan2020DSAC,ryg2020minmaxDSAC}, DDPG (Deep Deterministic Policy Gradient) \cite{lillicrap2015DDPG}, A3C (Asynchronous Advantage Actor-Critic) \cite{mnih2016asynchronous}, SAC (Soft Actor-Critic) \cite{Haarnoja2018SAC,Haarnoja2018ASAC}, TRPO (Trust Region Policy Optimization) \cite{schulman2015TRPO}, and PPO (Proximal Policy Optimization) \cite{schulman2017PPo}, have also been widely used to solve DT optimal control problems.

It should be pointed out that the state constraints were not considered in these ADP techniques mentioned above. For practical applications, however, most controlled systems must be subject to some state restrictions. Taking vehicle control in the path-tracking task as an example, in addition to considering the tracking performance, certain state functions of the vehicle must be constrained to the stability zone to prevent vehicle instability problems \cite{li2020predictive}. Compared with state constraints, input constraints can be easily confronted by introducing a nonquadratic cost function or directly constraining the output range of actor NN using some saturated functions, such as hyperbolic tangent function \cite{lewis2005definition,dong2016event,2015constrainedReinforcement}. To cope with the state constraints, most related ADP researches choose to transform the original problem into an unconstrained one by constructing additional system states or adding the state constraints to the objective function as a penalty \cite{2018outputconstrainedADP,2020outputconstrainedADP,2019stateconstrainedADP,SUN2018stateandinput_constrained}. However, the optimality of the original system may not be guaranteed since the controller of the transformed system has to spend extra efforts to ensure the satisfaction of the state constraints, which will certainly enlarge the cost function \cite{2018outputconstrainedADP}. Also, the trade-off between performance and state constraints may lead to constraint violations in some cases. Model predictive control (MPC) is a commonly used control method to solve control input online while satisfying a set of constraints \cite{borrelli2017MPC}. However, compared with ADP, complex systems such as non input-affine and nonlinear models are still big challenges for MPC. In addition, Achiam \emph{et al.} proposed a model-free RL algorithm, constrained policy optimization (CPO), to maximize rewards while enforcing constraints \cite{Achiam2017CPO}. But the constraint satisfaction of CPO, or its variant parallel CPO \cite{wenlu2020PCPO}, cannot ensure the policy safety, because its approximated constraint function is an expected and cumulative value.

Besides, existing ADP algorithms usually rely on hand-crafted features combined with linear value functions or policy representations, also called single NN in \cite{heydari2012single,zhang2012singleNN-ADP}. Obviously, the performance of these algorithms heavily relies on the quality of the feature representation. However, it is usually hard to design such features for high-dimensional nonlinear systems. Compared with single NNs, NNs with multiple hidden layers or deep NNs have the fitting ability to directly map the system states to optimal control inputs or value function without reliance on hand-crafted features \cite{lecun2015deep}. In addition, theoretical analysis and experimental results show that the multi-layer NN can usually converge to the nearly global minimum if it is over-parameterized \cite{allen2018convergence,du2018gradient}. This is why deep RL has outperformed traditional RL in many challenging domains, from games to robotic control \cite{sutton2018reinforcement}. Hence, in this paper, multi-layer NNs are employed to relax the need for hand-crafted features.

In this paper, we propose a new ADP algorithm, called constrained ADP (CADP), to solve optimal control problems with state constraints, which is applicable to general nonlinear systems with nonaffine saturated control inputs. The main contributions and advantages of this paper are summarized as follows:
\begin{enumerate}
    \item  A constrained generalized policy iteration (CGPI) framework is developed to handle state constraints, in which the traditional policy improvement process is transformed into a constrained policy optimization problem. Compared with most existing ADP algorithms \cite{chen2008generalizedHJB,al2008discrete,liu2015generalized} that use the traditional policy improvement, the proposed CADP method building on CGPI is applicable to optimal control problems with state constraints.
    \item For the approximated policy with high-dimensional parameters, directly solving the constrained policy optimization problem may be intractable due to the computational cost and the nonlinear characteristics of dynamics and the policy function. Therefore, some existing ADP researches consider state constraints by transforming the original problem into an unconstrained one \cite{2018outputconstrainedADP,2020outputconstrainedADP,2019stateconstrainedADP,SUN2018stateandinput_constrained}. However, the optimality may not be guaranteed since the controller of the transformed system has to spend extra efforts to ensure the satisfaction of the state constraints, which will enlarge the cost function \cite{2018outputconstrainedADP}. The proposed CADP algorithm deals with this situation by linearizing the constrained optimization problem locally into a quadratically constrained linear programming problem, and then obtains the optimal update of the policy function by solving its dual problem. Meanwhile, a trust region constraint is added to prevent excessive policy update, thus ensuring linearization accuracy. Besides, two recovery rules are proposed to update the policy in case that the primal problem is infeasible. 
    \item The proposed CADP algorithm employs the actor-critic architecture to approximate both policy and value functions by multi-layer NNs. According to the universal fitting ability and global convergence property of multi-layer NNs \cite{allen2018convergence,du2018gradient}, the value and policy networks of CADP directly map the system states to control inputs and value function, respectively. Therefore, compared with ADP researches \cite{chen2008generalizedHJB,2018outputconstrainedADP,heydari2012single,zhang2012singleNN-ADP,dierks2009NNmodel} that use single NN, CADP relaxes the need for hand-crafted features. Besides, different from \cite{chen2008generalizedHJB,2018outputconstrainedADP,2020outputconstrainedADP,2019stateconstrainedADP,SUN2018stateandinput_constrained,heydari2012single,zhang2012singleNN-ADP,dierks2009NNmodel,li2012ADPaffine-unknown} which are subject to input-affine systems since the policy needs to be analytically represented by the value function, CADP is applicable to arbitrary nonlinear nonaffine dynamics by optimizing the independent policy network.
\end{enumerate}

The paper is organized as follows. In Section \ref{sec.preliminary}, we provide the formulation of the DT optimal control problem, followed by the general description of GPI framework. Section \ref{sec.algorithm} presents the constrained ADP algorithm. In Section \ref{sec.simulation}, we present a simulation example that shows the effectiveness of the CADP algorithm for DT system. Section \ref{sec.conclusion} concludes this paper.

\section{Mathematical preliminaries}
\label{sec.preliminary}
\subsection{Notation}
For ease of presentation, we summarize the abbreviations and mathematical notations in Table \ref{tab.abbre} and Table \ref{tab.notation}, respectively.

\begin{table}[width=.95\linewidth,cols=2,pos=h]
\caption{Abbreviations}
\label{tab.abbre}
\begin{tabular}{ll}
\toprule
Abbreviation & Explanation  \\
\hline
ADP & Adaptive dynamic programming  \\
CADP & Constrained  ADP  \\
CGPI & Constrained generalized policy iteration \\
DT & Discrete-time  \\
GPI & Generalized policy iteration  \\
HJB & Hamilton-Jacobi-Bellman  \\
NN & Neural network  \\
PI&Policy iteration\\
\bottomrule
\end{tabular}
\end{table}

\begin{table}[width=.9\linewidth,cols=2,pos=h]
\caption{Mathematical Notations}
\label{tab.notation}
\begin{tabular}{ll}
\toprule
Symbol & Explanation  \\
\hline
$\mathbb{R}$ & the set of real numbers  \\
$\mathbb{N},\mathbb{N}_+$ & the set of natural numbers  \\
$\mathbb{R}^n$ & $n$-dimensional Euclidean space  \\
$k$ & time step index  \\
$x \in \mathbb{R}^n, u \in \mathbb{R}^m$ & state and control input vectors  \\
$f: \mathbb{R}^n \times \mathbb{R}^m \rightarrow \mathbb{R}^n$ & system function  \\
$\gamma\in(0,1)$ & discount factor \\
$\pi:\mathbb{R}^n \rightarrow \mathbb{R}^m$ & policy function  \\
$V:\mathbb{R}^n \rightarrow \mathbb{R} $ & value function  \\
$\theta \in \mathbb{R}^s, \omega \in \mathbb{R}^p$ & parameters vector of $\pi$ and $V$\\
$K$ & iteration index  \\
$\nabla_xF(\cdot)$ & the gradient of $F(\cdot)$ w.r.t. $x$  \\
$\|\cdot\|_2$&Euclidean norm\\
\bottomrule
\end{tabular}
\end{table}

\subsection{Discrete-time Hamilton-Jacobi-Bellman Equation}
Consider the discrete-time (DT) general time-invariant dynamical system 
\begin{equation}   
\label{eq.statefunction}
x_{k+1} = f(x_k,u_k),
\end{equation}
where $x_k \in \mathbb{R}^n$ and $u_k \in \mathbb{R}^m$ are the state vector and control input vector at time $k$, respectively, and $f:  \mathbb{R}^n \times \mathbb{R}^m \rightarrow \mathbb{R}^n$ is the system function. We assume that $f(x_k,u_k)$ is Lipschitz continuous on a compact set $\Omega$ that contains the origin, and that the system is stabilizable on $\Omega$, i.e., there exists a continuous policy $\pi(x)$, where $u_k=\pi(x_k)$, such that the system is asymptotically stable on $\Omega$. The system dynamics $f (x_k,u_k)$ is assumed to be known, which can be a nonlinear and input nonaffine analytic function only if $\frac{\partial f(x,u)}{\partial u}$ and $\frac{\partial f(x,u)}{\partial x}$ are available. The system input $u$ can be either constrained or unconstrained. Given the policy $\pi(x_k)$, we define its associated infinite-horizon value function as
\begin{equation}   
\label{eq.costfunction}
V^{\pi}(x_k) = \sum_{i=0}^{\infty} \gamma^{i} l(x_{k+i}, \pi(x_{k+i})),
\end{equation}
where $l(x_k,u_k):  \mathbb{R}^n \times \mathbb{R}^m \rightarrow \mathbb{R}$ is the utility function, and $\gamma \in (0,1)$ is the discount factor.

Furthermore, denoting the prediction horizon as $N$, \eqref{eq.costfunction} can be rewritten as 
\begin{equation}   
\label{eq.bellman}
V^{\pi}(x_k) = \sum_{i=0}^{N-1} \gamma^{i}l(x_{k+i},\pi(x_{k+i}))   +\gamma^N V^{\pi}(x_{k+N}),
\end{equation}
which is the well-known Bellman equation. Then the optimal control problem can now be formulated as finding a policy such that the value function associated with the system in \eqref{eq.statefunction} is minimized for $\forall x_k\in \Omega$. The minimized value function $V^*(x_k)$ defined by 
\begin{equation}
\label{eq.optimal_V}
V^*(x_k) = \min \limits_{\pi}V^{\pi}(x_k)
\end{equation}
satisfies the DT Hamilton-Jacobi-Bellman (HJB) equation or Bellman optimality equation
\begin{equation}   
\label{eq.bellman_optimality}
\begin{aligned}
V^{*}(x_k) =  \min \limits_{\pi}\Big\{\sum_{i=0}^{N-1} \gamma^{i}l(x_{k+i},\pi(x_{k+i})) + \gamma^N V^{*}(x_{k+N})\Big\}.
\end{aligned}
\end{equation}
Meanwhile, the optimal control $\pi^*(x_k)$ for $\forall x_k \in \Omega$ can be derived as
\begin{equation}   
\label{eq.optimalpolicy}
\pi^*(x_k) = \arg\min_{\pi} \Big\{\sum_{i=0}^{N-1} \gamma^{i}l(x_{k+i},\pi(x_{k+i})) + \gamma^N V^{*}(x_{k+N})\Big\}.
\end{equation}

To find the optimal control solution for the problem, one only needs to solve \eqref{eq.bellman_optimality} for the value function and then substitute the solution into \eqref{eq.optimalpolicy} to obtain the optimal control. However, due to the nonlinear nature of DT HJB, finding its solution is generally difficult or intractable. 

\subsection{Adaptive Dynamic Programming}
The proposed algorithm for DT optimal control problems used in this paper is motivated by generalized policy iteration (GPI) techniques \cite{sutton2018reinforcement}. GPI is an iterative method widely used in adaptive dynamic programming (ADP) and reinforcement learning (RL) algorithms to find the approximate solution of DT HJB. The ADP algorithms building on GPI usually employ actor-critic architecture to approximate both the value function and policy. In this study, both the value function and policy are approximated by multi-layer NNs, called value network (or critic network) $V(x; \omega)$ and policy network (or actor network) $\pi(x; \theta)$, where $\omega \in\mathbb{R}^p$ and $\theta\in\mathbb{R}^s$ are network parameters. These two networks directly build the map from the raw system states to the approximated value function and control inputs, respectively. In this case, no hand-crafted features or basis functions are needed. 

GPI involves two interacting processes: 1) policy evaluation, which drives the estimated value function towards the true value function for current policy based on \eqref{eq.bellman}, and 2) policy improvement, which improves the policy with respect to current estimated value function based on \eqref{eq.optimalpolicy}. 

Defining the accumulated future cost of state $x_k$ under policy $\pi_{\theta}$ and value $V_{\omega}$ as 
\begin{equation}
\label{eq.target_return}
G(x_k,\pi_{\theta},V_{\omega})=\sum_{i=0}^{N-1}\gamma^{i}l(x_{k+i},\pi(x_{k+i};\theta))+\gamma^N V(x_{k+N};\omega),
\end{equation}
the policy evaluation process of GPI proceeds by iteratively minimizing the following loss function:
\begin{equation}   
\label{eq.value_objective}
L(\omega) = \mathop{\mathbb{E}}_{x_k\sim d_x}\Big\{\frac{1}{2}\big(G(x_k,\pi_{\theta},V_{\omega})-V(x_{k};\omega)\big)^2\Big\},
\end{equation}
where $G(x_k,\pi_{\theta},V_{\omega})-V(x_{k};\omega)$ is usually called temporal difference (TD) error and $d_x$ denotes the state distribution over $x\in\Omega$.
Under the assumption of the universal approximation theorem of NNs \cite{Hornik1990Universal}, $d_x$ can be an arbitrary distribution as long as the probability density $p(x)>0$ for $\forall x \in \Omega$, such as uniform distribution. Therefore, the update gradient for the value network is given by
\begin{equation}   
\label{eq.value_update}
\nabla_{\omega}L(\omega) = \mathop{\mathbb{E}}_{x_k\sim d_x}\Big\{\big(V(x_{k};\omega)-G(x_k,\pi_{\theta},V_{\omega})\big)\nabla_{\omega}V(x_{k};\omega) \Big\}.
\end{equation}

In the policy improvement step, the parameters $\theta$ of the policy network are updated to minimize the objective function
\begin{equation}   
\label{eq.policy_objective}
J(\theta) = \mathop{\mathbb{E}}_{x_k\sim d_x}\Big\{G(x_k,\pi_{\theta},V_{\omega})\Big\}.
\end{equation}

Denoting the matrix $\frac{\partial x_{k+i}}{\partial \theta}\in\mathbb{R}^{s\times n}$ as $\phi_i$, $\frac{\partial u_{k+i}}{\partial \theta}\in\mathbb{R}^{s\times m}$ as $\psi_i$, the update gradient for the policy network is
\begin{equation}   
\label{eq.policy_update}
\begin{aligned}
&\nabla_{\theta}J(\theta)
\\&\ =\mathop{\mathbb{E}}_{x_k\sim d_x}\Big\{\sum_{i=0}^{N-1} \gamma^{i}\nabla_{\theta}l(x_{k+i},u_{k+i})+\gamma^{N}\nabla_{\theta}V(x_{k+N};\omega)\Big\}\\
&\ =\mathop{\mathbb{E}}_{x_k\sim d_x}\Big\{\sum_{i=0}^{N-1} \gamma^{i}\Big[{\phi_{i}}\frac{\partial l(x_{k+i},u_{k+i})}{\partial x_{k+i}}+{\psi_{i}}\frac{\partial l(x_{k+i},u_{k+i})}{\partial u_{k+i}}\Big] \\
&\qquad \qquad \qquad \quad \quad \quad \quad \quad \quad +\gamma^N \phi_{N}\frac{\partial V(x_{k+N};\omega)}{ \partial x_{k+N}}\Big\},
\end{aligned}
\end{equation}
where
\begin{equation}
\nonumber
\phi_{i} =\phi_{i-1}\frac{\partial f(x_{k+i-1},u_{k+i-1})}{\partial x_{k+i-1}} + \psi_{i-1}\frac{\partial f(x_{k+i-1},u_{k+i-1})}{\partial u_{k+i-1}},
\end{equation} 
with $\phi_0=0$, and
\begin{equation}
\nonumber
\psi_{i} =\phi_{i}\frac{\partial\pi(x_{k+i};\theta)}{\partial x_{k+i}} + \nabla_{\theta}\pi(x_{k+i};\theta).
\end{equation}

In practice, $\nabla_{\omega}L(\omega)$ and $\nabla_{\theta}J(\theta)$ are usually approximated by the sample average. Any off-the-shelf  NN  optimization  methods  can  be used  to update the value and policy networks, including stochastic gradient descent (SGD),  RMSProp,  Adam \cite{ruder2016optimizationmethod}. Taking the  SGD  method as an example, the pseudo-code of GPI can be summarized as Algorithm \ref{alg:GPI}. Algorithm \ref{alg:GPI} will iteratively converge to the optimal control policy $\pi(x_k;\theta^*)=\pi^*(x_k)$ and  value function $V(x_k;\omega^*)=V^*(x_k)$ for $\forall x_k \in \Omega$.  Proofs  of convergence and optimality have been given in \cite{liu2017ADP,sutton2018reinforcement}.
 
\begin{algorithm}[!htb]
\caption{GPI Framework}
\label{alg:GPI}
\begin{algorithmic}
\STATE Initial with arbitrary $\theta_0$, $\omega_0$, learning rates $\alpha_c$ and $\alpha_a$
\STATE Initialize iteration index $K=0$
\REPEAT
\STATE Rollout $N$ steps from $x_k\in\Omega$ with policy $\pi_{\theta_K}$\\
\STATE Receive and store $x_{k+i}$, $i\in[1,N]$
\STATE Policy evaluation: 
\STATE \ Calculate $G(x_k,\pi_{\theta_K},V_{\omega_K})$, $\nabla_{\omega_K}L(\omega_K)$ using \eqref{eq.target_return}, \eqref{eq.value_update}
\STATE \ Update value function using \begin{equation}
\label{eq.updaterule_V}
\omega_{K+1} = -\alpha_c \nabla_{\omega_K}L(\omega_K)+\omega_K 
\end{equation}
\STATE Policy improvement: 
\STATE \ Calculate $G(x_k,\pi_{\theta_K},V_{\omega_{K+1}})$, $\nabla_{\theta_K}J(\theta_K)$ using \eqref{eq.target_return}, \eqref{eq.policy_update}
\STATE \ Update policy using \begin{equation}
\label{eq.updaterule_P}
\theta_{K+1} = -\alpha_a \nabla_{\theta_K}J(\theta_K) + \theta_K 
\end{equation}
$K=K+1$
\UNTIL Convergence  
\end{algorithmic}
\end{algorithm}

\section{Constrained ADP}
\label{sec.algorithm}
\subsection{Constrained Policy Improvement}
One drawback of the policy update rule in \eqref{eq.updaterule_P} is that it is not suitable for optimal control problems with state constraints. However, for practical applications, most controlled systems must be subject to some state restrictions, such as vehicles \cite{li2020predictive}, wave energy converters \cite{2019stateconstrainedADP} and robots \cite{li2015robotictrajectory}. Although the state constraints can be added to the objective function as a penalty, it is often difficult to balance the constraint requirements with the control objectives. The optimality may not be guaranteed since the controller has to spend extra efforts to ensure the satisfaction of the state constraints. Besides, the trade-off between performance and state constraints may lead to constraint violations in some cases. 

In this paper, the state constraints of future $N$ steps are introduced to transfer the policy improvement process into a constrained optimization problem. Assuming there are $\tau_{\text{max}}$ kinds of state constraints, 
the $\tau$th state constraint can be formulated as:
\begin{equation}   
\label{eq.state_constraint}
J_{C_\tau}(x_{k+i+1}) \le b_{\tau}, \quad i\in[0,N-1],
\end{equation}
where $J_{C_\tau}(x_{k+i+1}):\mathbb{R}^n \rightarrow \mathbb{R}$ is the $\tau$th state function bounded above by boundary $b_{\tau}$. Therefore, the policy improvement process can be transformed into the following constrained optimization problem:
\begin{equation}   
\label{eq.constrained_PI}
\begin{aligned}
\theta_{K+1}&=\arg\min_\theta  J(\theta) \\
s.t. \quad & x_{k+i+1} = f(x_{k+i},\pi(x_{k+i};\theta)),& i\in[0,N-1],\\
&  J_{C_\tau}(x_{k+i+1}) \le b_{\tau},  & \tau\in[1,\tau_{\text{max}}].
\end{aligned}
\end{equation}
There are a total of $M=N\times\tau_{\text{max}}$ state constraints in \eqref{eq.constrained_PI}. In this paper, we refer to \eqref{eq.constrained_PI} as the constrained policy improvement. Then we develop the constrained generalized policy iteration (CGPI) framework, which builds on GPI by replacing the policy improvement process in Algorithm \ref{alg:GPI} with \eqref{eq.constrained_PI}. The pseudo-code of CGPI is shown in Algorithm \ref{alg:CGPI}.

\begin{algorithm}[!htb]
\caption{CGPI Framework}
\label{alg:CGPI}
\begin{algorithmic}
\STATE Initial with arbitrary $\theta_0$, $\omega_0$, learning rates $\alpha_c$
\STATE Initialize iteration index $K=0$
\REPEAT
\STATE Rollout $N$ steps from $x_k\in\Omega$ with policy $\pi_{\theta_K}$\\
\STATE Receive and store $x_{k+i}$, $i\in[1,N]$
\STATE Policy evaluation: 
\STATE \ Calculate $G(x_k,\pi_{\theta_K},V_{\omega_K})$, $\nabla_{\omega_K}L(\omega_K)$ using \eqref{eq.target_return}, \eqref{eq.value_update}
\STATE \ Update value function using \eqref{eq.updaterule_V}
\STATE Constrained policy improvement: 
\STATE \ Construct $J(\theta)$ under $V_{\omega_{K+1}}$ using \eqref{eq.policy_objective}
\STATE \ Update policy using \eqref{eq.constrained_PI}
\STATE $K=K+1$
\UNTIL Convergence  
\end{algorithmic}
\end{algorithm}

In Appendix \ref{sec.appendix_CPI_tabular}, we prove the convergence and global optimality of a policy iteration variant of Algorithm \ref{alg:CGPI} based on tabular setting. Besides, as described in Appendix \ref{sec.appendix_CPI_function} and \ref{sec.appendix_CGPI}, the convergence results can be further extended to the case of function approximation and Algorithm \ref{alg:CGPI}.

\begin{remark}
The state constraints need to be reasonable to ensure \eqref{eq.constrained_PI} is feasible. For practical applications, the state constraints usually come from the physical limitations or boundaries of controlled systems \cite{li2020predictive,2019stateconstrainedADP,li2015robotictrajectory}. Taking vehicle control in the path-tracking task as an example, in addition to considering the tracking performance, certain state functions of the vehicle must be constrained to the stability zone to prevent vehicle instability problems \cite{li2020predictive}. 
\end{remark}

\subsection{Approximate Solution}
For policies with high-dimensional parameter spaces $\theta \in \mathbb{R}^s$, directly solving \eqref{eq.constrained_PI} may be intractable due to the computational cost and the nonlinear characteristics of NNs and dynamics. Local linearization is an effective trick to deal with this situation \cite{schulman2015TRPO,Achiam2017CPO,wenlu2020PCPO}. Firstly, we can linearize the objective function and state constraints at the $K$th iteration around current policy $\pi(x_k;\theta_K)$ using Taylor's expansion theorem. To ensure the approximation accuracy,
policy $\pi(x_k;\theta_{K+1})$ must be in a small neighborhood of $\pi(x_k;\theta_{K})$. This means that we need to add a policy constraint to \eqref{eq.constrained_PI} to avoid excessive policy update.

Inspired by \cite{schulman2015TRPO}, one effective way to limit the policy change is to constrain the difference between the new policy $\pi(x_k;\theta_{K+1})$ and the old policy $\pi(x_k;\theta_{K})$. Firstly, we define the following function
\begin{equation}
\nonumber
D_{\pi}(\theta; \theta_K) \doteq \mathbb{E}_{x_k\sim d_x}[\|\pi(x_k;\theta)-\pi(x_k;\theta_K)\|^2_2]
\end{equation}
to measure the difference between $\pi(x_k;\theta)$ and $\pi(x_k;\theta_{K})$. Then the following policy constraint, also known as the trust region constraint, can be constructed:
\begin{equation}   
\label{eq.trust_region}
D_{\pi}(\theta; \theta_K)  \le \delta,
\end{equation}
where $\delta\in\mathbb{R}_+$ is the trust region boundary. In this case, the policy update step is positively correlated with $\delta$. Then, \eqref{eq.constrained_PI} can be adapted to a trust region version:
\begin{equation}   
\label{eq.primal}
\begin{aligned}
\theta_{K+1}&=\arg\min_\theta  J(\theta) \\
s.t. \quad & x_{k+i+1} = f(x_{k+i},\pi(x_{k+i};\theta)),& i\in[0,N-1],\\
&  J_{C_\tau}(x_{k+i+1}) \le b_{\tau},  & \tau\in[1,\tau_{\text{max}}], \\
&  D_{\pi}(\theta; \theta_K)  \le \delta.
\end{aligned}
\end{equation}

For a small step size $\delta$, the objective function $J(\theta)$ and state functions $J_{C_\tau}$ in the $K$th iteration can be well-approximated by linearizing around current policy $\pi(\theta_K)$ using Taylor's expansion theorem. Denoting $\Delta \theta = \theta - \theta_K$ and $J_{C_{\tau,i}}=J_{C_\tau}(x_{k+i+1})$, it follows that:
\begin{equation}
\nonumber
J(\theta) \approx J(\theta_K)+(\nabla_{\theta}J(\theta)\big|_{\theta=\theta_K})^\top{\Delta \theta},
\end{equation}
and
\begin{equation}
\nonumber
J_{C_\tau}(x_{k+i+1}) \approx J_{C_\tau}(x_{k+i+1})|_{\theta=\theta_K}+\big(\nabla_{\theta}J_{C_{\tau,i}}\big|_{\theta=\theta_K}\big)^\top{\Delta \theta},
\end{equation}
where
\begin{equation}
\nonumber
\nabla_{\theta}J_{C_{\tau,i}} ={\phi_{i+1}\frac{\partial J_{C_\tau}(x_{k+i+1})}{\partial x_{k+i+1}}}, \quad i\in[0,N-1].
\end{equation} 
In addition, since $D_{\pi}(\theta;\theta_K)$ and its gradient are both zero at $\theta=\theta_K$, the trust region constraint is well-approximated by second-order Taylor expansion:
\begin{equation}
\nonumber
\begin{aligned}
&D_{\pi}(\theta; \theta_K) 
\\&\ \approx D_{\pi}(\theta_K; \theta_K) +\big(\nabla_{\theta}D_{\pi}(\theta; \theta_K)\big|_{\theta=\theta_K}\big)^\top\Delta\theta+ \frac{1}{2} {\Delta \theta}^{\top}H{\Delta \theta}\\
&\ =\frac{1}{2} {\Delta \theta}^{\top}H{\Delta \theta},
\end{aligned}
\end{equation} 
where $H\in\mathbb{R}^{s\times s}$ is the Hessian of $D_{\pi}$ with respect to $\theta$, i.e., $H_{i,j}=\frac{\partial^2D_{\pi}(\theta; \theta_K) }{\partial\theta_i\partial\theta_j}\Big|_{\theta=\theta_K}$. Since $D_{\pi}(\theta; \theta_K)\ge0$ for $\forall \theta \in \mathbb{R}^s$, $H$ is always positive semi-definite. In keeping with other work in the literature \cite{Achiam2017CPO,wenlu2020PCPO}, we will assume it to be positive-definite in the following.

Denoting $g=\nabla_{\theta}J/{\|\nabla_{\theta}J\|}_2$, $c_{j}=\nabla_{\theta}J_{C_{\tau,i}}/{\|\nabla_{\theta}J_{C_{\tau,i}}\|}_2$, and $z_{j} = (J_{C_{\tau,i}}|_{\theta=\theta_K} - b_\tau)/{\|\nabla_{\theta}J_{C_{\tau,i}}\|}_2$, where $j=(\tau-1)\times N+i+1 \in [1,M]$. With $C \doteq [c_1, c_2, ... , c_M]\in\mathbb{R}^{s\times M}$ and $z \doteq [z_1, z_2, ... , z_M]^{\top}$,  the approximation to \eqref{eq.primal} is:
\begin{equation}   
\label{eq.primal_linear}
\begin{aligned}
\min_{\Delta \theta}  &  \quad g^{\top}{\Delta \theta}\\
s.t. &  \quad  z + C^{\top} {\Delta \theta} \le 0, \\ &  \quad \frac{1}{2} {\Delta \theta}^{\top}H{\Delta \theta} \le \delta.
\end{aligned}
\end{equation}
Denoting the optimal solution of \eqref{eq.primal_linear} as $\Delta \theta^*$, the update rule for the constrained policy improvement process is
\begin{equation}
\label{eq.updaterule}
\nonumber
\theta_{K+1} = \Delta \theta^* + \theta_K.
\end{equation}

Although \eqref{eq.primal_linear} is a convex constrained optimization problem,  directly solving it will take lots of computation time and resources because the dimension of variables $\Delta \theta \in \mathbb{R}^s$ is very large (usually over 10 thousand). Since \eqref{eq.primal_linear} is convex, it can also be solved using the dual method when feasible. The Lagrange function of \eqref{eq.primal_linear} can be expressed as
\begin{equation}   
\label{eq.Lagrange_2}
\nonumber
L_a({\Delta \theta}, \lambda, \nu) =  g^{\top}{\Delta \theta} + \lambda (\frac{1}{2} {\Delta \theta}^{\top}H{\Delta \theta}-\delta) + \nu^{\top}(z + C^{\top} {\Delta \theta}),
\end{equation}
where $\lambda\in\mathbb{R}$ and $\nu \in \mathbb R^M$.
Then the dual to \eqref{eq.primal_linear} is
\begin{equation}   
\label{eq.dual}
\max_{\lambda\ge0,\nu\ge0}  \min_{\Delta \theta}\  L_a({\Delta \theta}, \lambda, \nu).
\end{equation}
The gradient of $L_a$ with respect to parameters ${\Delta \theta}$ can be calculated as
\begin{equation}   
\label{eq.delta_Lagrange}
\nonumber
\nabla_{{\Delta \theta}} L_a({\Delta \theta}, \lambda, \nu) =  g + \lambda H{\Delta \theta} +  C\nu.
\end{equation}
When $\nabla_{{\Delta \theta}} L_a({\Delta \theta}, \lambda, \nu)=0$, we have
\begin{equation}   
\label{eq.theta}
{\Delta \theta} = - \frac{H^{-1}(g+C\nu)}{\lambda},\quad \lambda > 0.
\end{equation}
By taking \eqref{eq.theta} into \eqref{eq.dual}, the dual to \eqref{eq.primal_linear} can be expressed as
\begin{equation}   
\label{eq.dual_2}
\max_{\lambda>0,\nu\ge0}  -\frac{1}{2\lambda}(\mu+\nu^{\top}S\nu+2\nu^{\top}r)-\lambda\delta+\nu^{\top}z,
\end{equation}
where $\mu=g^{\top}H^{-1}g$, $S=C^{\top}H^{-1}C$, $r=C^{\top}H^{-1}g$.
Let
\begin{equation}   
\label{eq.L_optimal}
\nonumber
L(\lambda, \nu) \doteq  -\frac{1}{2\lambda}(\mu+\nu^{\top}S\nu+2\nu^{\top}r)-\lambda\delta+\nu^{\top}z,
\end{equation}
then we can rewrite \eqref{eq.dual_2} with 
\begin{equation}   
\label{eq.dual_3}
\min_{\lambda>0,\nu\ge0} \ -L(\lambda, \nu).
\end{equation}

Problem \eqref{eq.dual_3} is a bound-constrained convex optimization problem with only $M+1$ variables, which is also equal to the number of constraints in \eqref{eq.primal} but much smaller than the dimension of $\Delta \theta \in \mathbb{R}^s$. Therefore, compared with \eqref{eq.primal}, the optimal solution of \eqref{eq.dual_3} can be  solved more easily and efficiently by using off-the-shelf methods such as L-BFGS-B and truncated Newton method \cite{zhu1997BFGS,nocedal2006optimization}. Supposing $\lambda^*$, $\nu^*$ are the optimal solutions to \eqref{eq.dual_3}, $\theta_{K+1}$ can be updated as
\begin{equation}   
\label{eq.update_rule_2}
\theta_{K+1} = -\frac{H^{-1}(g+C\nu^*)}{\lambda^*} + \theta_{K}.
\end{equation}

Note that for a high-dimensional policy $\pi(x_k;\theta)$, it is prohibitively costly to compute $H$ and invert $H^{-1}$, which poses a huge challenge for computing $H^{-1}g$ and $H^{-1}C$ in \eqref{eq.dual_3}. In this paper, we approximately compute them using the conjugate gradient method without reliance on forming the full matrix $H$ or $H^{-1}$ \cite{schulman2015TRPO,Achiam2017CPO}.

\subsection{Feasibility}
On the one hand, the initial policy $\pi_{\theta_{0}}$ may be infeasible. On the other hand, due to the approximation errors induced by linearization, the optimal solution $\Delta \theta^*$ of \eqref{eq.primal_linear} at the $K$th iteration may be a bad update, and then a new policy $\pi_{\theta_{K+1}}$ that fails to satisfy state constraints may be produced. This may cause the optimization problem \eqref{eq.primal_linear} of the $K+1$th iteration to be infeasible. In other words, the feasible region of \eqref{eq.primal_linear} may be empty in some cases, i.e., $\Theta_A \cap \Theta_B = \varnothing$,
where $\Theta_A =\{\Delta \theta: z + C^{\top} {\Delta \theta} \le 0 \}$, and $\Theta_B =\{\Delta \theta: \frac{1}{2} {\Delta \theta}^{\top}H{\Delta \theta} \le \delta \}$. 

Hence, before solving the dual problem \eqref{eq.dual_3}, we check whether the feasible region is empty by calculating the minimum trust region boundary, which makes the trust region intersect with the intersection of half-planes of all linear constraints:
\begin{equation}   
\label{eq.mintrust_region}
\begin{aligned}
\min_{\Delta \theta}  &  \quad \frac{1}{2} {\Delta \theta}^{\top}H{\Delta \theta}\\
s.t. &  \quad  z + C^{\top} {\Delta \theta} \le 0. 
\end{aligned}
\end{equation}
Denoting the optimal solution to \eqref{eq.mintrust_region} as $\Delta \theta_{\text{min}}$, then the minimum trust region boundary that makes \eqref{eq.primal_linear} feasible is $\delta_{\text{min}} = \frac{1}{2} {\Delta \theta_{\text{min}}}^{\top}H{\Delta \theta_{\text{min}}}$, and it is clear that
\begin{equation}
\label{eq.judgerule}
\left\{
\begin{aligned}
  &\Theta_A \cap \Theta_B = \varnothing,  \quad \delta_{\text{min}} > \delta,\\
  &\Theta_A \cap \Theta_B \ne \varnothing, \quad \delta_{\text{min}} \le \delta.
\end{aligned}
\right.
\end{equation} 
The value $\delta_{\text{min}}$ can be efficiently obtained by solving the following dual problem:
\begin{equation}   
\label{eq.dual_trustregion}
\max_{\nu\ge0}  \ -\frac{\nu^{\top}S\nu}{2}+\nu^{\top}z.
\end{equation}
Suppose $\nu^{\dagger}$ is the optimal solution of \eqref{eq.dual_trustregion}, then
\begin{equation}
\nonumber    
\delta_{\text{min}}=-\frac{{\nu^{\dagger}}^{\top}S\nu^{\dagger}}{2}+{\nu^{\dagger}}^{\top}z.
\end{equation}

It is known from \eqref{eq.judgerule} that the magnitude of the value $\delta$ directly affects the feasibility of \eqref{eq.primal_linear}. Denoting the expected trust region boundary as $\delta_a$, if $\delta_{\text{min}} \le \delta_a$, we can directly solve \eqref{eq.dual_3} with $\delta = \delta_a$. For the infeasible case, i.e., $\delta_{\text{min}} > \delta_a$, a recovery method is needed to calculate a reasonable policy update. By introducing the  recovery trust region boundary $\delta_b$, which is slightly greater than $\delta_a$, we propose two recovery rules according to the value of $\delta_{\text{min}}$: 1) If $\delta_b \ge\delta_{\text{min}} > \delta_a$, we solve \eqref{eq.dual_3} with $\delta=\delta_b$ for $\lambda^*$ and $\nu^*$; 2) If $\delta_{\text{min}}> \delta_b$, we recover the policy by adding the state constraints as a penalty to the original objective function:
\begin{equation}   
\label{eq.penalty}
\begin{aligned}
\min_{\Delta \theta}  &  \quad \Big((1-\eta)g +\eta\sum_{j=1}^{M} \alpha_j {c_j}\Big)^{\top} {\Delta \theta} \\
s.t. &  \quad \frac{1}{2} {\Delta \theta}^{\top}H{\Delta \theta} \le \delta_b,
\end{aligned}
\end{equation}
where $\eta$ is the hyper-parameter that trades off the importance between the original objective function and the penalty term, $\alpha_j$ is the weight of the state constraint corresponding to $c_j$, which is calculated by: 
\begin{equation}   
\label{eq.weight}
\nonumber
\alpha_j = \frac{p_j e^{z_j}}{\sum_{j=1}^{M}p_je^{z_j}},
\end{equation}
where $p_{j}=5$ if $z_{j}>0$, $p_{j}=1$ otherwise, which penalizes violations of the corresponding state constraint. Defining $g_p \doteq (1-\eta)g +\eta\sum_{j=1}^{M} \alpha_j {c_j}$, the dual to \eqref{eq.penalty} can be expressed as:
\begin{equation}   
\label{eq.dual_penalty}
\max_{\lambda>0}  \  -\frac{\mu_p}{2\lambda}-\lambda\delta_b,
\end{equation}
where $\mu_p={g_p}^{\top}H^{-1}g_p$. In this case, we can easily find the policy recovery rule:
\begin{equation}   
\label{eq.update_rule_recover}
\theta_{K+1} = - \sqrt{\frac{2\delta_b}{\mu_p}}H^{-1}g_p + \theta_{K}.
\end{equation}

\subsection{Constrained Adaptive Dynamic Programming}
\begin{figure}[!htb]
\centering{\includegraphics[width=0.48\textwidth]{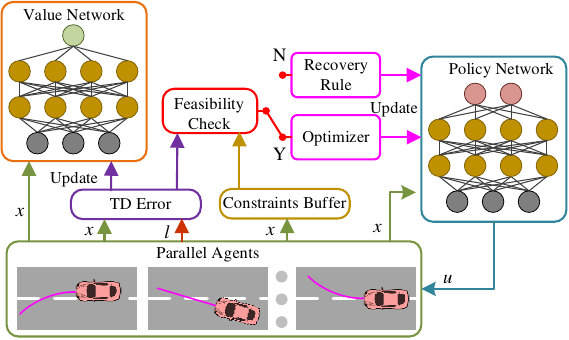}}
\caption{CADP diagram. The value function and policy are approximated by two NNs, called value network and policy network, respectively. The value network is updated by minimizing loss function \eqref{eq.value_objective}. If problem \eqref{eq.primal_linear} is feasible, we update the policy with its optimal solution; otherwise, we update with the recovery rules. Multiple parallel agents are employed to explore different parts of the state space. All the state constraints of these parallel agents are stored in the constraints buffer. For each iteration, we only consider $M$ state constraints randomly selected from the constraints buffer.} \label{f:CDADP}
\end{figure}

By introducing the trust region constraint, approximation solution method, and recovery rules, we adapt Algorithm \ref{alg:CGPI}
to Algorithm \ref{alg:CDADP}, called constrained adaptive dynamic programming (CADP). Besides, we also incorporate the parallel exploring trick widely used in RL \cite{mnih2016asynchronous,duan2020HRL,duan2020DSAC} to accelerate training and improve stability (See Fig. \ref{f:CDADP}). In particular, we use multiple parallel agents to explore different state spaces, thereby removing correlations in the training set. All the state constraints of these parallel agents are stored in the constraints buffer. Due to the computational burden caused by estimating the matrices $C$, $S$ and solving \eqref{eq.dual_3}, the speed of the policy optimization process will decrease as the number of state constraints increases. For each iteration, we only consider $M$ state constraints randomly selected from the constraints buffer. 

\begin{algorithm}[!htb]
\caption{CADP Algorithm}
\label{alg:CDADP}
\begin{algorithmic}
\STATE Initial with arbitrary $\theta_0$, $\omega_0$, learning rates $\alpha_c$
\STATE Initialize iteration index $K=0$
\REPEAT
\STATE Rollout $N$ steps from $x_k\in\Omega$ with policy $\pi_{\theta_K}$\\
\STATE Receive and store $x_{k+i}$, $i\in[1,N]$
\STATE Policy evaluation: 
\STATE \ \ Calculate $G(x_k,\pi_{\theta_K},V_{\omega_K})$, $\nabla_{\omega_K}L(\omega_K)$ using \eqref{eq.target_return}, \eqref{eq.value_update}
\STATE \ \ Update value function using \eqref{eq.updaterule_V}
\STATE Constrained policy improvement: 

\setlength{\leftskip}{0.8em}
\STATE Solve dual problem \eqref{eq.dual_trustregion} for $\delta_{\text{min}}$ 
\IF{$\delta_{\text{min}}\le\delta_b$}

\setlength{\leftskip}{1.2em}
\STATE $
\begin{aligned}
\delta = \left\{
\begin{aligned}
  &\delta_a,  &\quad \delta_{\text{min}}\le\delta_a\\
  &\delta_b, &\quad\text{else}
\end{aligned}
\right.
\end{aligned}$
\STATE Solve dual problem \eqref{eq.dual_3} for $\lambda^*, \nu^*$ \\
\STATE Update policy with \eqref{eq.update_rule_2}

\setlength{\leftskip}{0.8em}
\ELSE

\setlength{\leftskip}{1.2em}
\STATE Update policy with \eqref{eq.update_rule_recover}

\setlength{\leftskip}{1.2em}
\ENDIF

\setlength{\leftskip}{0.0em}
\STATE $K=K+1$
\UNTIL Convergence  
\end{algorithmic}
\end{algorithm}

\begin{remark}
The performance of ADP approaches proposed in \cite{chen2008generalizedHJB,2018outputconstrainedADP,heydari2012single,zhang2012singleNN-ADP,dierks2009NNmodel} that use single NN heavily relies on the quality of hand-crafted features. It is usually hard to design such features for high-dimensional nonlinear systems. Nevertheless, the update rules \eqref{eq.updaterule_V} and \eqref{eq.update_rule_2} of the proposed CADP algorithm are applicable to most approximate functions, such as multi-layer NNs. Hence, according to Lemma \ref{lemma.approximation} and \ref{lemma.global_min}, by employing multi-layer NNs to represent the policy and value function, CADP can directly learn a map from system states to control inputs and value function, thereby relaxing the need for hand-crafted features.
\end{remark}

\begin{remark}
The proposed CADP method relies on the knowledge of the system dynamics $f(x,u)$. For controlled systems with unknown dynamics, we can use supervised learning to learn an approximated model from data, such as an NN-based model. In recent years, many algorithms based on the learned NN-dynamics have been proposed \cite{nagabandi2018MBMF,heess2015NNmodel}. Similarly, given a real system with unknown dynamics, we can first learn an NN-based model, and then apply the CADP algorithm to find the nearly optimal policy. 
\end{remark}

\begin{remark}
For the proposed CADP algorithm, according to \eqref{eq.value_update} and \eqref{eq.primal_linear}, the system $f(x,u)$ can be an arbitrary analytic function only if it is differentiable, i.e., $\frac{\partial f(x,u)}{\partial u}$ and $\frac{\partial f(x,u)}{\partial x}$ are available. Therefore, different from \cite{chen2008generalizedHJB,2018outputconstrainedADP,2020outputconstrainedADP,2019stateconstrainedADP,SUN2018stateandinput_constrained,heydari2012single,zhang2012singleNN-ADP,dierks2009NNmodel,li2012ADPaffine-unknown} that are subject to input-affine systems since the policy needs to be analytically represented by the value function, CADP is applicable to arbitrary nonlinear systems with nonaffine saturated inputs. Nonlinear nonaffine systems are very common in practical applications, such as vehicle dynamics \cite{li2020predictive} and NN-based models learned from data \cite{nagabandi2018MBMF,heess2015NNmodel}.
\end{remark}

\begin{remark}
Given a practical optimal control problem with known dynamics, we only need to formulate the policy optimization process as \eqref{eq.primal}. Then, CADP can be directly used to find the nearly optimal policy without reliance on hand-crafted features. The learned offline policy maps the states to the corresponding nearly optimal control inputs, which can be directly applied to the controlled system to realize online control. 
\end{remark}

\section{Simulation}
\label{sec.simulation}
\subsection{Problem Description}
To evaluate the performance of the CADP algorithm, we choose the vehicle lateral and longitudinal control in the path-tracking task as an example. It is a nonlinear and nonaffine system control problem with state constraints \cite{li2020predictive}. The control objective is to maximize the vehicle speed, while maintaining a small tracking error and ensuring that the vehicle stays within the stability region. The system states and control inputs of this problem are listed in Table \ref{tab.state}, and the vehicle parameters are listed in Table \ref{tab.parameters}. In keeping with other studies \cite{li2020predictive,li2017sharecontrol,kong2015kinematic}, we assume the states in Table \ref{tab.state} are observable. Note that the system frequency used for simulation is different from the sampling frequency $f$. The vehicle is controlled by a saturated actuator, where  $\dot{\xi} \in [-0.35, 0.35]$ and $\dot{a}\in [-2, 2]$. According to the continuous-time vehicle dynamics given in \cite{li2020predictive,li2017sharecontrol,kong2015kinematic}, the corresponding discrete-time dynamics can be obtained using forward Euler methods \cite{li2017sharecontrol,kong2015kinematic}, which can be described by:
\begin{equation}
\nonumber
x = 
\begin{bmatrix}
  v_y \\
  r \\
  v_x \\{}
  \phi \\
  y \\
  \xi\\
  a
\end{bmatrix}
,u = 
\begin{bmatrix}
  \dot{\xi}\\
  \dot{a}
\end{bmatrix}  
,x_{k+1}=
\begin{bmatrix}
  \frac{F_{y\rm{f}}\cos\xi + F_{y\rm{r}}}{m} - v_x r\\
  \frac{d_{\rm{f}}F_{y\rm{f}}\cos\xi - d_{\rm{r}}F_{y\rm{r}}}{I_{z}}\\
  a + v_y r \\
  r\\
  v_x \sin\phi + v_y \cos\phi \\
  \dot{\xi}\\
  \dot{a}
\end{bmatrix}\frac{1}{f} + x_k,
\end{equation} 
where $F_{y\rm{f}}$ and $F_{y\rm{r}}$ are the lateral tire forces of the front and rear tires respectively. The lateral tire forces are approximated according to the Fiala tire model:
\begin{equation}
\nonumber
\begin{split}
&F_{y\dagger}  = -\mathrm{sgn}(\alpha_\dagger) * \min   \\
& \Big\{ \left|C_\dagger\tan\alpha_\dagger\Big(\frac{C_\dagger^2(\tan\alpha_\dagger)^2}{27(\mu_\dagger F_{z\dagger})^2} -  \frac{{C_\dagger}\left |\tan\alpha_\dagger \right |}{3\mu_\dagger F_{z\dagger}} + 1 \Big)\right|,\left|\mu_\dagger F_{z\dagger}\right| \Big\}, 
\end {split}
\end{equation}
where $\alpha_{\dagger}$ is the tire slip angle, $F_{z\dagger}$ is the tire load, $\mu_{\dagger}$ is the lateral friction coefficient, and the subscript $\dagger \in \{\rm{f},\rm{r}\}$ represents the front or rear tires. The slip angles can be calculated from the geometric relationship between the front/rear axle and the  center of gravity (CG): 
\begin{equation}
\nonumber
\alpha_{\rm{f}} = \arctan (\frac{v_y+d_{\rm{f}}r}{v_x})-\xi, \quad \alpha_{\rm{r}} = \arctan (\frac{v_y-d_{\rm{r}}r}{v_x}).
\end{equation} 
Let $\alpha_{\text{max},\dagger}$ represent the tire slip angle when the tire fully-sliding behavior occurs, calculated as:
\begin{equation}
\nonumber
\alpha_{\text{max},\dagger} =  \frac{3\mu_{\dagger} F_{z\dagger}}{C_{\dagger}}.
\end{equation} 

Assuming that the rolling resistance is negligible, the lateral friction coefficient of the front/rear wheel is:
\begin{equation}
\nonumber
\mu_{\dagger} = \frac{\sqrt{(\mu F_{z\dagger})^2-(F_{x\dagger})^2}}{F_{z\dagger}},
\end{equation} 
where $F_{x\rm{f}}$ and $F_{x\rm{r}}$ are the longitudinal tire forces of the front and rear tires respectively, calculated as:
\begin{equation}
\nonumber
\{F_{x\rm{f}},F_{x\rm{r}}\} = \left\{
\begin{aligned}
  &\{0 ,ma\},  &\quad a\ge0,\\
  &\{\frac{ma}{2}, \frac{ma}{2}\}, &\quad a<0.
\end{aligned}
\right.
\end{equation} 

The loads on the front and rear tires can be approximated by: 
\begin{equation}
\nonumber
F_{z\rm{f}} = \frac{d_{\rm{r}}}{d_{\rm{f}}+d_{\rm{r}}}mg, \quad F_{z\rm{r}} = \frac{d_{\rm{f}}}{d_{\rm{f}}+d_{\rm{r}}}mg.
\end{equation} 

\begin{table}
\centering
\caption{State and Control Input}
\label{tab.state}
\begin{tabular}{lll}
\toprule
Explanation &Symbol & Unit  \\
\hline
Lateral velocity at center of gravity (CG)&$v_y$ & [m/s]  \\
Yaw rate &$r$ & [rad/s] \\
Longitudinal velocity at CG&$v_x$ & [m/s] \\
Yaw angle between vehicle \& trajectory &$\phi$ & [rad] \\
Distance between CG 
\& trajectory&$y$ & [m] \\
Front wheel angle &$\xi$ & [rad]  \\
Longitudinal acceleration &$a$ & [m/$\mathrm{s}^2$] \\
\hline

 Rate of change of $a$ &$\dot{a}$ & [m/$\mathrm{s}^3$] \\
 Rate of change of  $\xi$&$\dot{\xi}$ & [rad/s]  \\
\bottomrule
\end{tabular}
\end{table}

\begin{table}
\centering
\caption{Vehicle Parameters}
\label{tab.parameters}
\begin{tabular}{lll}
\toprule
Explanation &Symbol & Value  \\
\hline
Front wheel cornering stiffness &$C_{\rm{f}}$ & -88000 [N/rad]  \\
Rear wheel cornering stiffness &$C_{\rm{r}}$ & -94000 [N/rad] \\
Distance from CG to front axle &$d_{\rm{f}}$ & 1.14 [m] \\
Distance from CG to rear axle &$d_{\rm{r}}$ & 1.40 [m] \\
Mass &$m$ & 1500 [kg] \\
Polar moment of inertia at CG &$I_z$ & 2420 [kg$\cdot\mathrm{m}^2$] \\
Tire-road friction coefficient&$\mu$ & 1.0 \\ 
Sampling frequency &$f$ & 40 [Hz] \\ 
Simulation frequency & & 200 [Hz] \\ 
\bottomrule
\end{tabular}
\end{table}

To ensure vehicle stability, the yaw rate $r$ at the CG and the slip angles $\alpha_{\dagger}$ 
should be subject to the following constraints:
\begin{equation}
\label{eq.vehicle-constraint}
\begin{aligned}
-r_{\text{max}} \le &r  \le r_{\text{max}},\\
-\alpha_{\text{max},\rm{f}} \le& \alpha_{\rm{f}} \le \alpha_{\text{max},\rm{f}},\\
-\alpha_{\text{max},\rm{r}} \le& \alpha_{\rm{r}} \le \alpha_{\text{max},\rm{r}},\\
\end{aligned}
\end{equation} 
where $r_{\text{max}}=\frac{\mu_{\rm{r}} g}{v_x}$.

The utility function is
\begin{equation}
\nonumber
\begin{aligned}
&l(x,u) = \\
&\frac{2(v_x-30)^2+80y^2 +40r^2+100(2\phi^2+\xi^2+{\dot{\xi}}^2)+a^2+{\dot{a}}^2}{2000}.
\end{aligned}
\end{equation} 
Hence, the policy optimization problem of this example is given by
\begin{equation}
\nonumber
\begin{aligned}
\min_\theta   &\mathop{\mathbb{E}}_{x_k\sim d_x}\Big\{\sum_{i=0}^{N-1} \gamma^{i}l(x_{k+i},u_{k+i})   + V(x_{k+N};\omega)\Big\}\\
s.t. \quad    &x_{k+i+1} = f(x_{k+i},\pi(x_{k+i};\theta)), \quad i\in[0,N-1],\\
 &{\Big|\frac{r v_x}{ \mu_{\rm{r}}}\Big|}_{k+i+1} \le g,\\
&\Big|\frac{\alpha_{\rm{f}}}{ \mu_f}\Big|_{k+i+1}  \le \frac{3 F_{z\rm{f}}}{ C_{\rm{f}}},\\
&\Big|\frac{\alpha_{\rm{r}}}{ \mu_{\rm{r}}}\Big|_{k+i+1}  \le \frac{3 F_{z\rm{r}}}{ C_{\rm{r}}},\\
  &D_{\pi}(\theta; \theta_K)  \le \delta,
\end{aligned}
\end{equation}
where ${\Big|\frac{r v_x}{ \mu_{\rm{r}}}\Big|}_{k+i+1}$, $\Big|\frac{\alpha_{\rm{f}}}{ \mu_f}\Big|_{k+i+1}$ and $\Big|\frac{\alpha_{\rm{r}}}{ \mu_{\rm{r}}}\Big|_{k+i+1}$ are the state constraint functions of state $x_{k+i+1}$ bounded above by $g$, $\frac{3 F_{z\rm{f}}}{ C_{\rm{f}}}$ and $\frac{3 F_{z\rm{r}}}{ C_{\rm{r}}}$ respectively. It is clear that the form of this problem is the same as \eqref{eq.primal}, which means that we can directly train the vehicle control policy using the proposed CADP algorithm.

\subsection{Algorithm Details}
In this paper, the value function and policy are represented by fully-connected NNs, which have the same architecture except for the output layers. For each network, the input layer is composed of the states, followed by 5 hidden layers using exponential linear units (ELUs) as activation functions, with 32 units per layer. The output of the value network is a linear unit, while the output layer of the policy network is set as a $tanh$ layer with two units, multiplied by the matrix $[0.35, 2]$ to confront bounded controls. We use Adam method to update the value network $V(x;\omega)$ with the learning rate of $8\times10^{-4}$. Other hyper-parameters of this problem are shown in Table \ref{tab.hyperparameters}.

\begin{table}[width=.9\linewidth,cols=2,pos=h]
\centering
\caption{Hyper-parameters}
\label{tab.hyperparameters}
\begin{tabular}{lll}
\toprule
Parameters &Symbol&Value\\
\hline
agent number &  & 256  \\
prediction horizon & $N$ & 30 \\
number of state constraints & $M$ & 10 \\
discount factor &$\gamma$ & 0.98  \\
trust region boundary &$\delta_a$ & $0.003^3$ \\
recovery trust region boundary &$\delta_b$ & $0.006^3$ \\
penalty factor &$\eta$ & 0.8 \\
\bottomrule
\end{tabular}
\end{table}

We compare the CADP algorithm with four other algorithms, namely GPI, TRADP (i.e., GPI with trust region constraint), penalty TRADP (P-TRADP, i.e., update policy network by directly solving \eqref{eq.penalty} with $\eta=0.2, 0.4, 0.6$ respectively), and CPO (constrained policy optimization \cite{Achiam2017CPO}). Note that TRADP can be considered as a special case of P-TRADP, in which $\eta=0$. The value $\eta$ plays a different role in CADP and P-TRADP algorithms. In CADP , $\eta$ in \eqref{eq.penalty} works only when $\delta_{\rm{min}}> \delta_b$. Therefore, we can take a relatively large $\eta$ to make \eqref{eq.primal_linear} feasible as soon as possible.

\subsection{Result Analysis}

\begin{figure}
\centering{\includegraphics[width=0.48\textwidth]{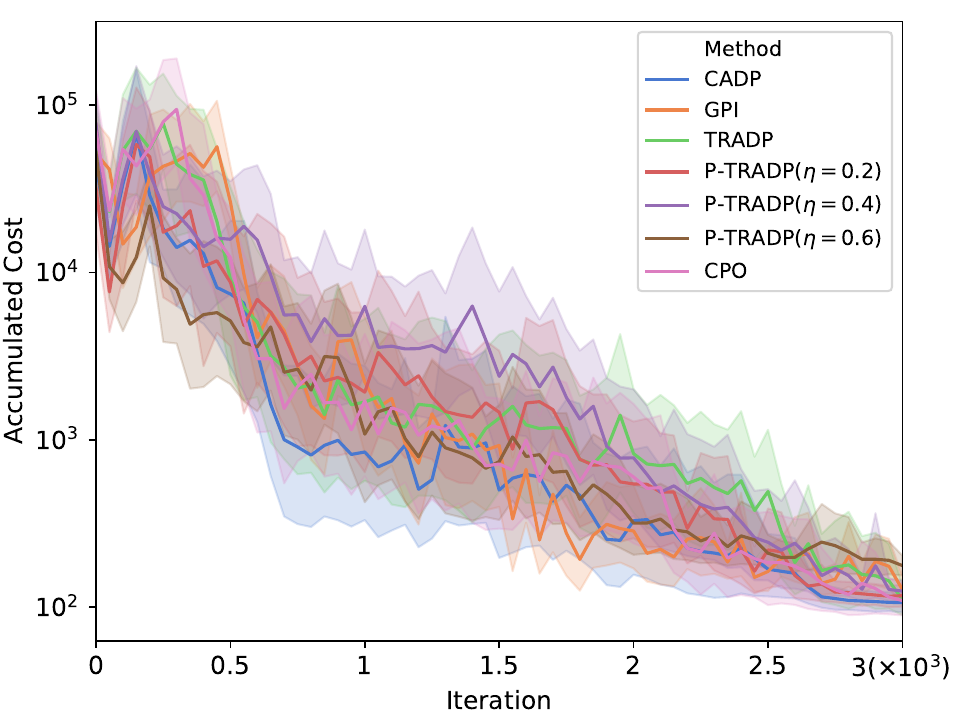}}
\caption{Training performance. Solid lines are average values over 20 runs. Shaded regions correspond to $95\%$ confidence interval.}\label{f:training}
\end{figure}

\begin{figure*}
\centering
\captionsetup[subfigure]{justification=centering}
\subfloat[\label{subFig:policyperformance}]{\includegraphics[width=0.4\textwidth]{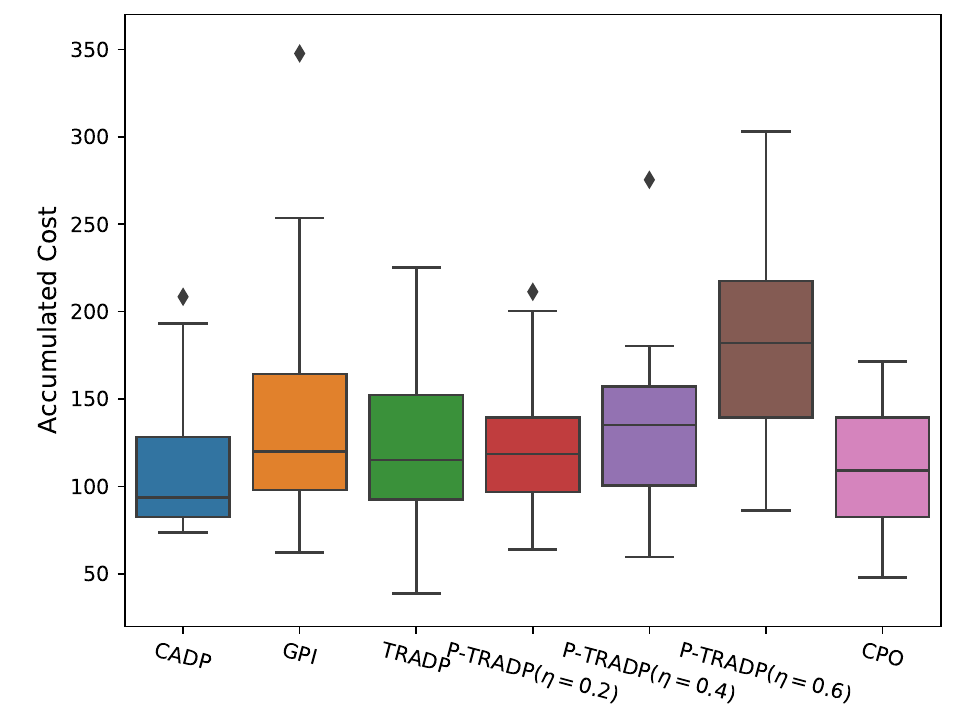}} 
\subfloat[ \label{subFig:r}]{\includegraphics[width=0.4\textwidth]{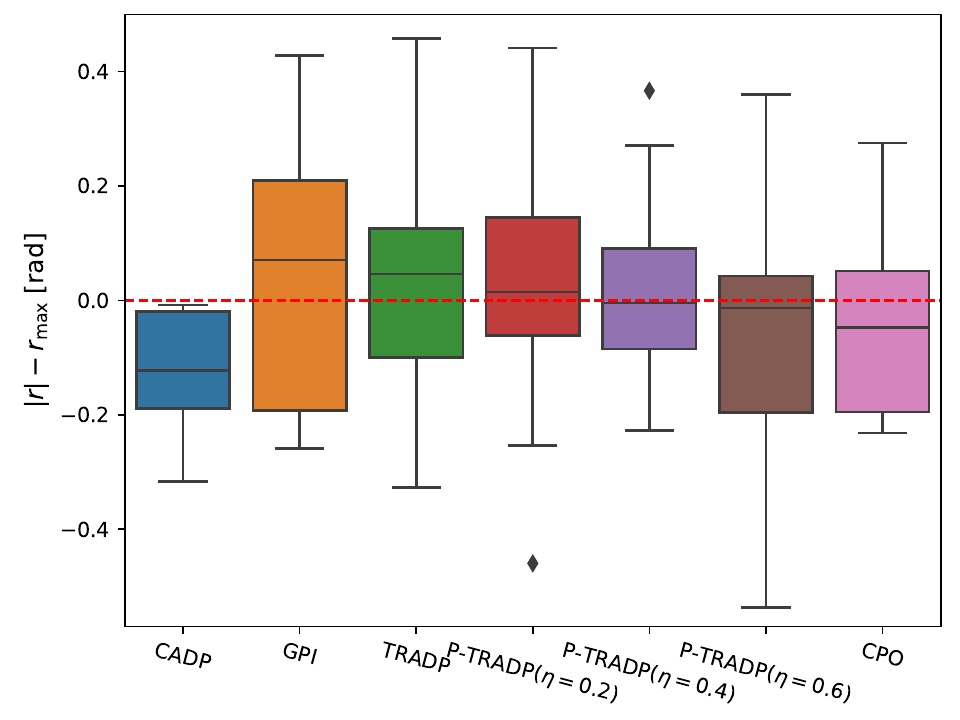}} \\
\subfloat[\label{subFig:ar}]{\includegraphics[width=0.4\textwidth]{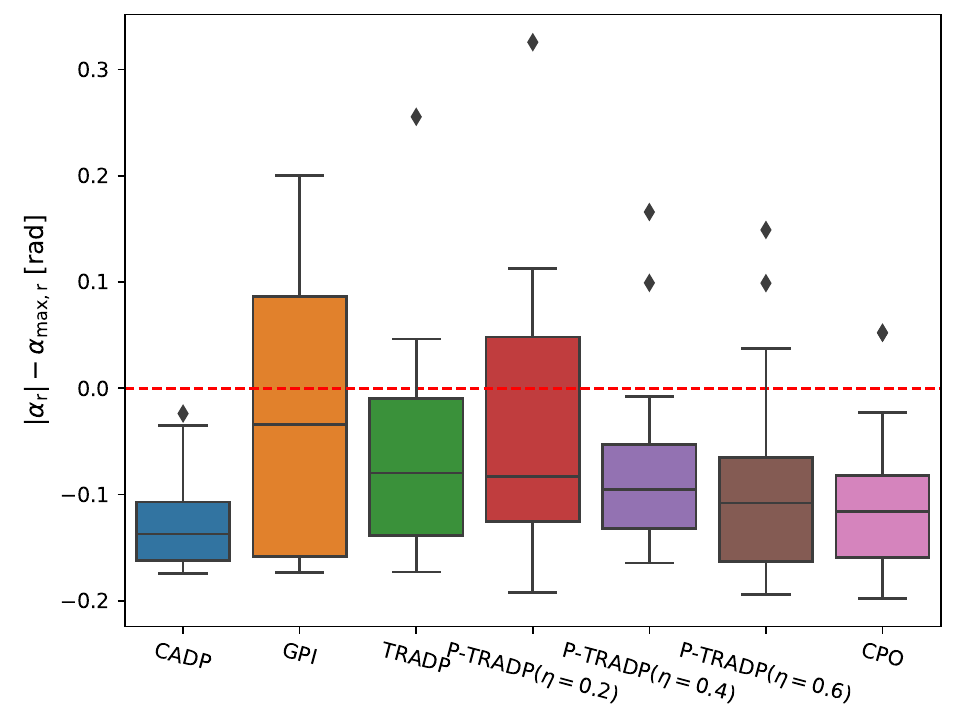}} 
\subfloat[\label{subFig:af}]{\includegraphics[width=0.4\textwidth]{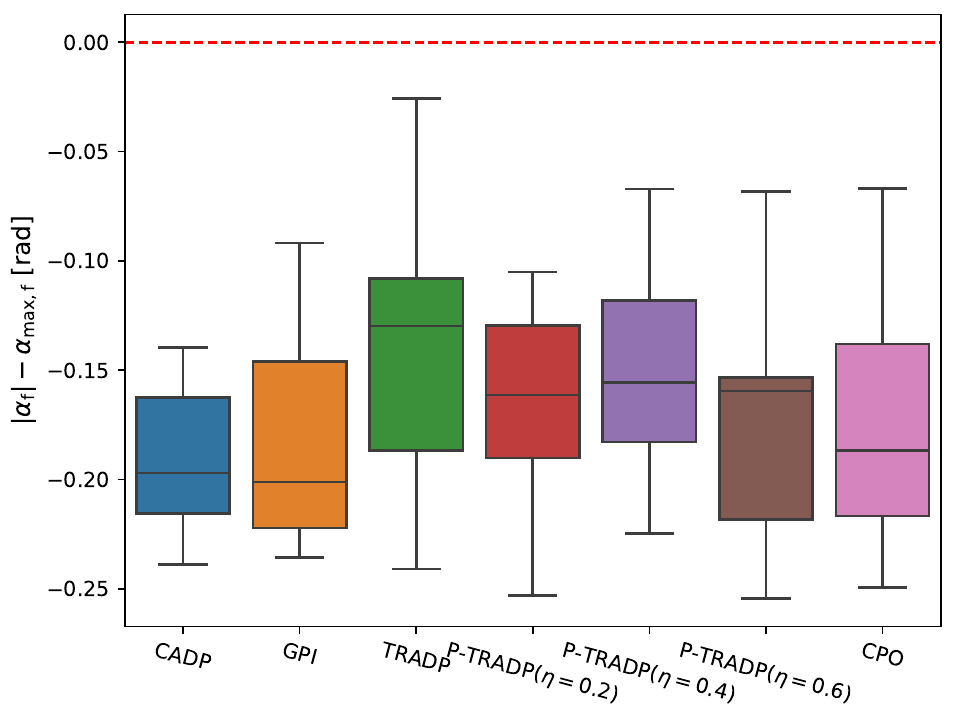}} 
\caption{Algorithm comparison. Each box plot is drawn based on values of 20 runs. Values greater than 0 (marked as red dashed lines) indicate violations of the corresponding state constraints. (a) Policy performance. (b) Maximum value of $|r|-r_{\text{max}}$ per simulation. (c) Maximum value of $|\alpha_{\rm{r}}|-\alpha_{\text{max,r}}$ per simulation. (d) Maximum value of $|\alpha_{\rm{f}}|-\alpha_{\text{max,f}}$ per simulation.}
\label{f:comparison}
\end{figure*}

\begin{figure*}
\centering
\captionsetup[subfigure]{justification=centering}
\subfloat[\label{subFig:state_trajectory_GPI}]{\includegraphics[width=0.31\textwidth]{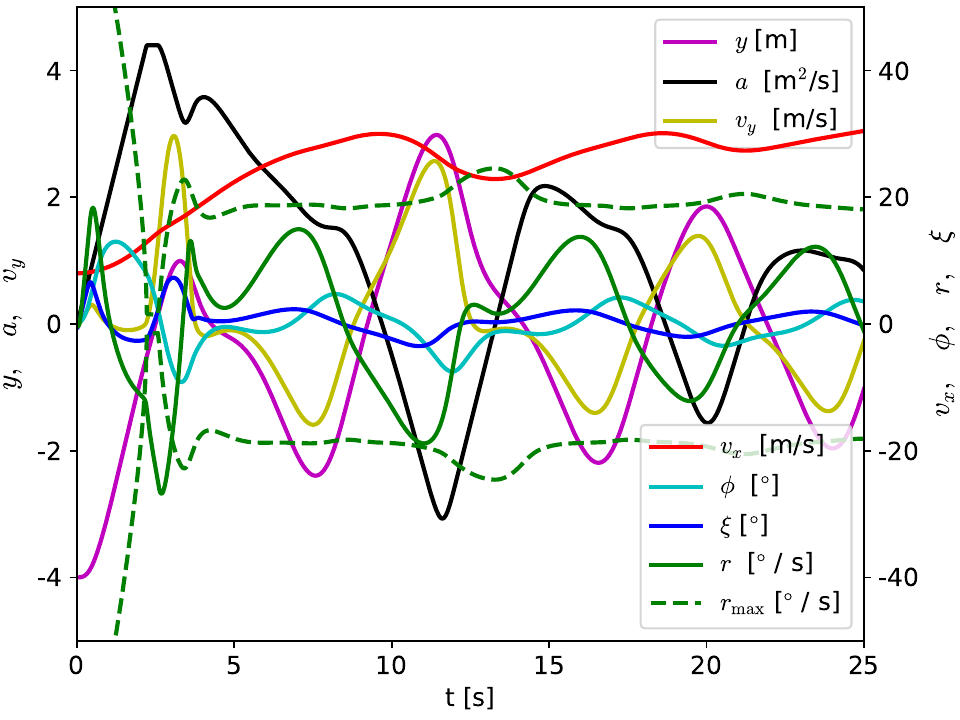}} \quad
\subfloat[\label{subFig:state_trajectory_penalty}]{\includegraphics[width=0.31\textwidth]{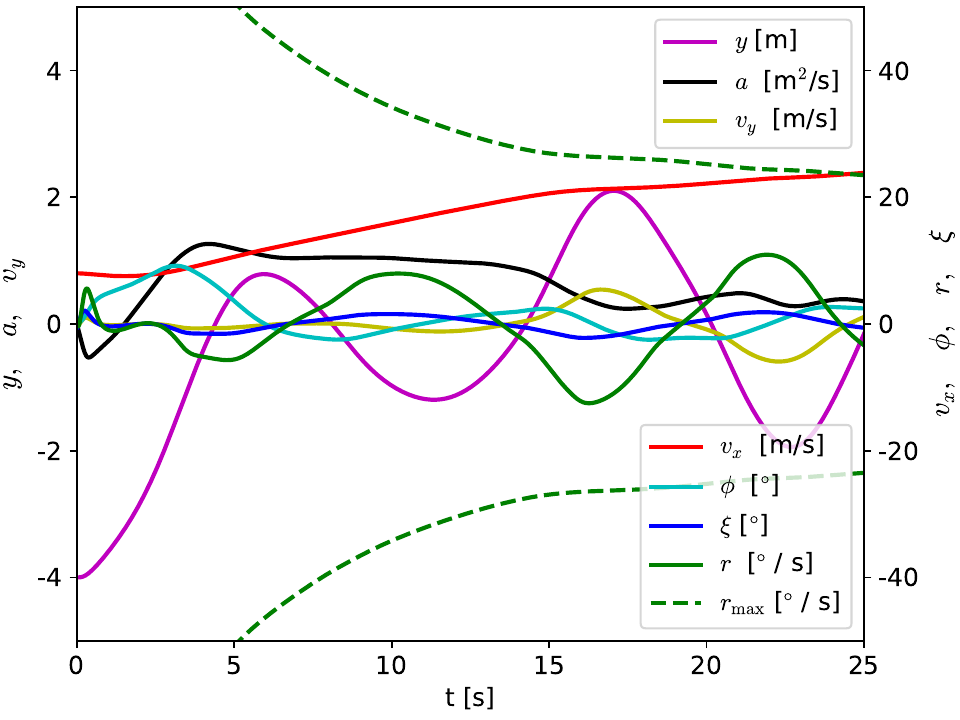}}  \quad
\subfloat[\label{subFig:state_trajectory_CDADP}]{\includegraphics[width=0.31\textwidth]{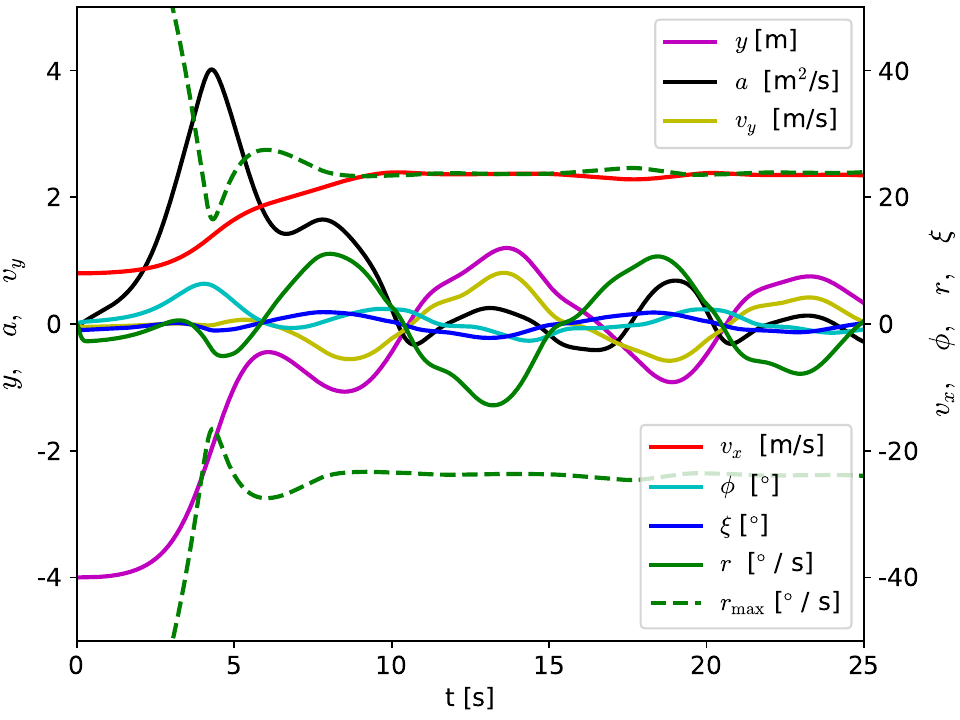}} \\
\subfloat[\label{subFig:constraint_GPI}]{\includegraphics[width=0.31\textwidth]{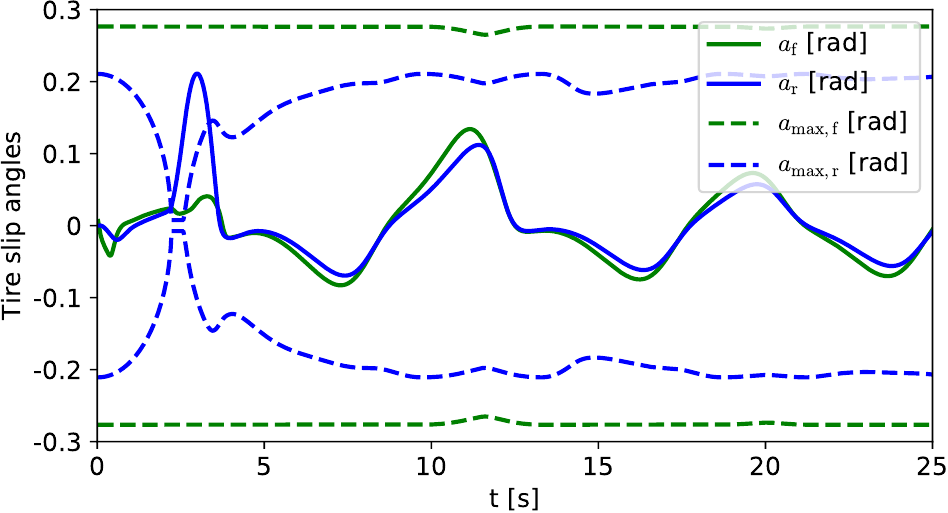}} \quad
\subfloat[\label{subFig:constraint_penalty}]{\includegraphics[width=0.31\textwidth]{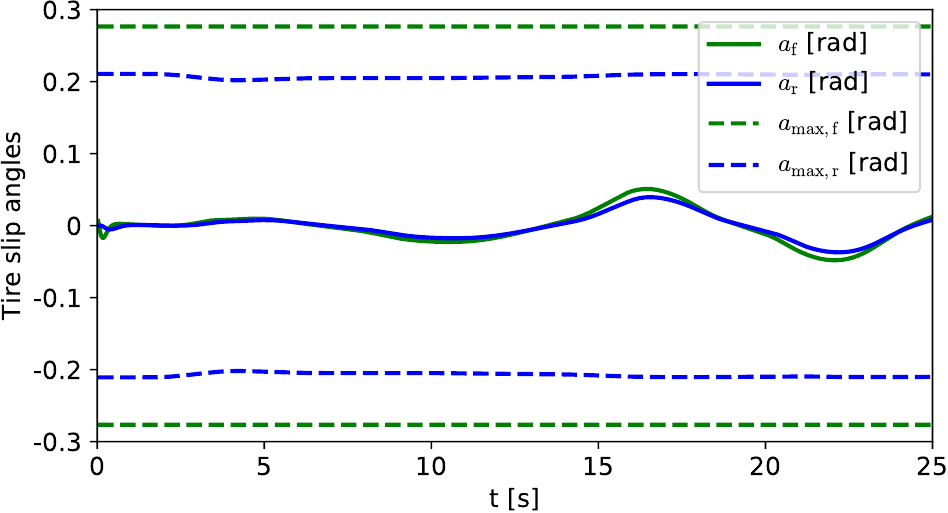}}  \quad
\subfloat[\label{subFig:constraint_CDADP}]{\includegraphics[width=0.31\textwidth]{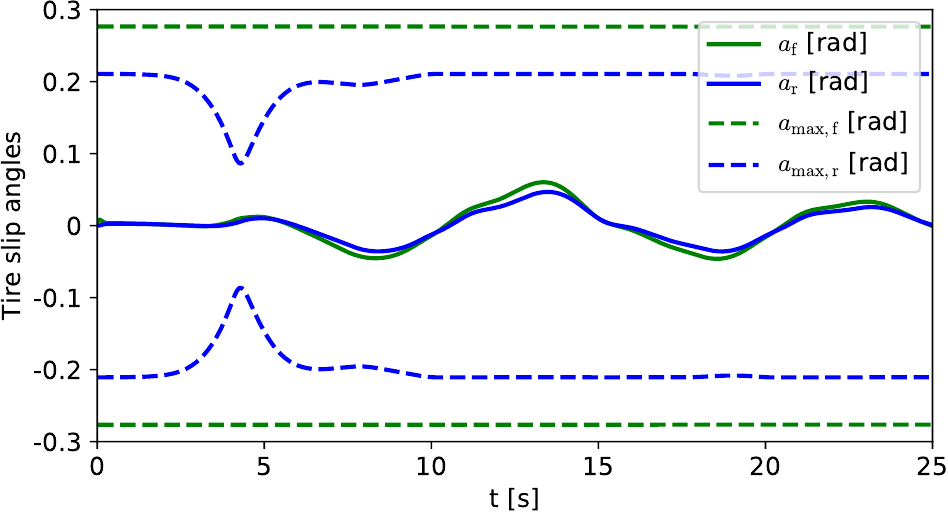}} \\
\subfloat[\label{subFig:trajectory_GPI}]{\includegraphics[width=0.31\textwidth]{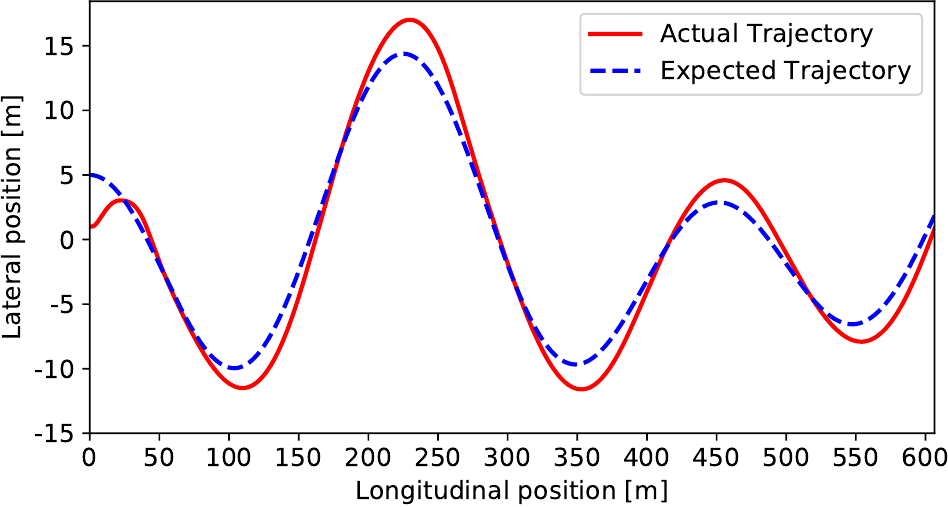}} \quad
\subfloat[\label{subFig:trajectory_penalty}]{\includegraphics[width=0.31\textwidth]{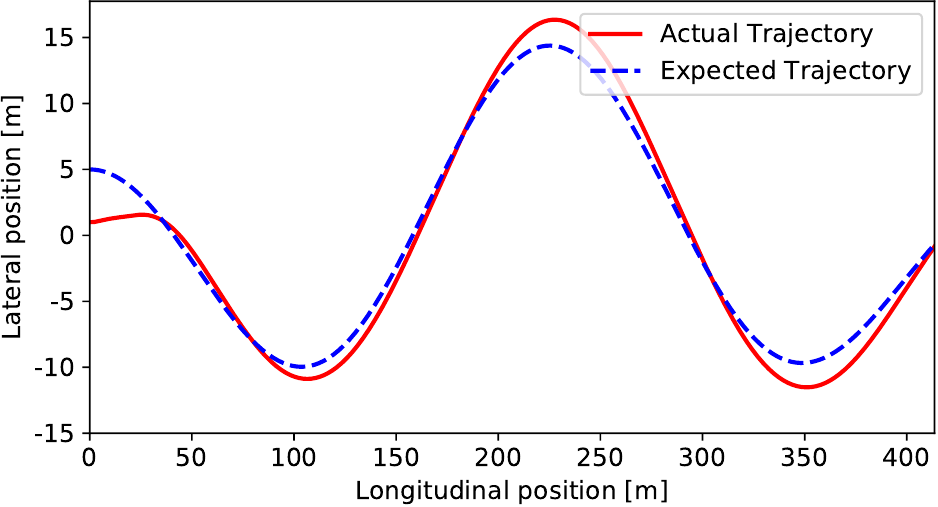}}  \quad
\subfloat[\label{subFig:trajectory_CDADP}]{\includegraphics[width=0.31\textwidth]{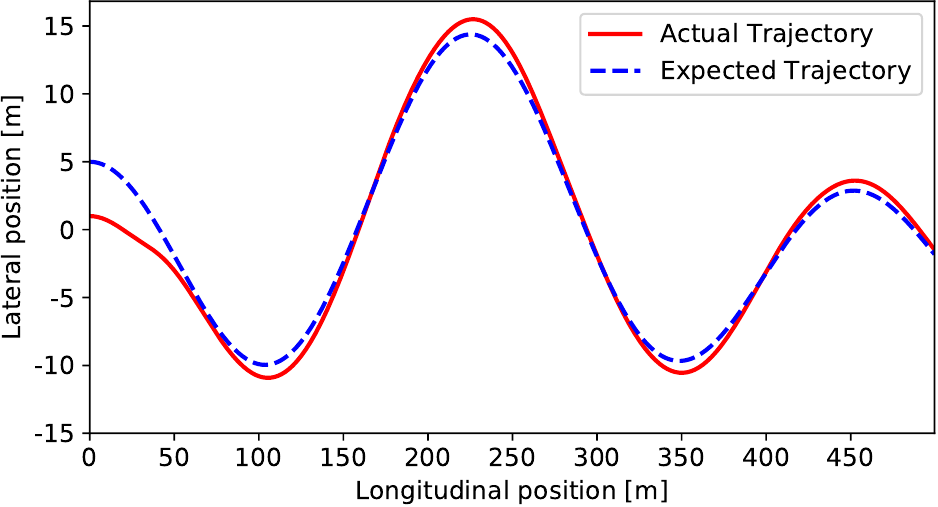}} \\
\caption{Simulation results. (a) State curves of GPI. (b) State curves of P-TRADP ($\eta=0.6$). (c) State curves of CADP. (d) Tire slip angles curves of GPI. (e) Tire slip angles curves of P-TRADP ($\eta=0.6$). (f) Tire slip angles curves of CADP. (g) Trajectory of GPI. (h) Trajectory of P-TRADP ($\eta=0.6$). (i) Trajectory of CADP.}
\label{f:simulation}
\end{figure*}

We train 20 different runs of each algorithm with different random seeds, with evaluations every 100 iterations. Each evaluation measures the policy performance by calculating the undiscounted accumulated cost function of 1000 steps (25s) during the simulation period starting from a random initialized state. The learning curves are shown in Fig. \ref{f:training}. Basically, all algorithms converge within 3000 iterations.

Fig. \ref{f:comparison} compares the performance and constraint satisfaction of each algorithm. As shown in Fig. \ref{subFig:policyperformance}, CADP matches or outperforms all other baseline ADP algorithms in policy performance. Fig. \ref{subFig:r}, \ref{subFig:ar} and \ref{subFig:af} show the maximum value of $|r|-r_{\text{max}}$, $|\alpha_{\rm{r}}|-\alpha_{\text{max,r}}$, and $|\alpha_{\rm{f}}|-\alpha_{\text{max,f}}$ for each simulation, respectively. Take Fig. \ref{subFig:r} as an example, from \eqref{eq.vehicle-constraint}, $|r|-r_{\text{max}}>0$ indicates the corresponding state constraint is violated. Results show that only CADP meets all three state constraints during the simulation. For TRADP and P-TRADP algorithms, the cumulative cost increases with the penalty factor $\eta$. This is because the learned policy needs to balance the cost term and constraint term. In particular, when $\eta=0.6$, the accumulated cost of the learned policy is almost twice that of CADP because the constraint term accounts for too much of the objective function in \eqref{eq.penalty}. Even so, for the P-TRADP algorithm with $\eta=0.6$, the situation in which the controlled system violates the constraints still exists. CPO considers the state constraint by limiting the expected discount cumulative state function, which is approximated by an additional NN. The transformed constraints of CPO are not equivalent to the original constraints given in \eqref{eq.state_constraint}, so the satisfaction of state constraints are still not guaranteed. CADP deals with the state constraints by solving \eqref{eq.primal}, which enables it to find the optimal policy from all feasible policies that satisfy the state constraints.

The typical simulation results by applying GPI, P-TRADP ($\eta=0.6$) and CADP are presented in Fig. \ref{f:simulation}. As shown in Fig. \ref{subFig:state_trajectory_GPI}, \ref{subFig:constraint_GPI} and \ref{subFig:trajectory_GPI}, the policy learned by GPI is too aggressive, resulting in large acceleration $a$, front wheel angle $\xi$ and velocity $v_x$, thus violating state constraints $r_{\rm max}$ and $\alpha_{\rm max, r}$. In contrast, the policy learned by P-TRADP ($\eta=0.6$) is conservative in the control of $a$ and $\xi$, leading to large accumulated cost (see Fig. \ref{subFig:state_trajectory_penalty}, \ref{subFig:constraint_penalty} and \ref{subFig:trajectory_penalty}). As a comparison, the policy learned by CADP has a good balance between tracking performance and constraint satisfaction. This example demonstrates the effectiveness of CADP in solving the nonaffine nonlinear optimal control problems with multiple state constraints.

\section{Conclusions}
\label{sec.conclusion}

This paper proposes a constrained adaptive dynamic programming (CADP) algorithm to solve optimal control problems with multiple state constraints. The proposed algorithm is applicable to general nonlinear systems with nonaffine and saturated control inputs. Firstly, a constrained generalized policy iteration (CGPI) framework is developed to handle state constraints by transforming the traditional policy improvement process into a constrained policy optimization problem. The proposed CADP algorithm is an actor-critic variant of CGPI, in which both policy and value functions are approximated by multi-layer NNs to directly map the system states to control inputs and value function respectively, which relaxes the need for hand-crafted features. Due to the computational cost and the nonlinear characteristics of NNs, it may be intractable to directly solve the constrained policy optimization problem. For calculation convenience, the proposed CADP linearizes the constrained optimization problem locally into a quadratically constrained linear programming problem, and then obtains the optimal update of the policy network by solving its dual problem. Meanwhile, a trust region constraint is added to prevent excessive policy update, thus ensuring linearization accuracy. We determine the feasibility of the policy optimization problem by calculating the minimum trust region boundary and update the policy using two recovery rules when infeasible.


We apply CADP and six other baseline algorithms to the vehicle control problem in the path-tracking task. Results show that CADP matches or outperforms all other baseline algorithms in policy performance without violating any state constraints during the simulation. In the future, we will extend the proposed algorithm to uncertain systems by combining other techniques, such as robust control \cite{liu2013robust} and fuzzy adaptive control \cite{sun2020Event-Fuzzy,sun2020Fuzzy}.

\appendix
\section{Appendix}
\subsection{Convergence Proofs of Constrained Policy Iteration in the Tabular Case}
\label{sec.appendix_CPI_tabular}
We refer to the combination of policy iteration (PI), a special case of GPI, and constrained policy improvement \eqref{eq.constrained_PI} as constrained policy iteration (CPI). We first discuss the convergence and optimality of CPI. In a version of CPI for a tabular setting, we maintain two tables to estimate policy $\pi$ and the corresponding value function $V^{\pi}$ respectively. The pseudo-code of CPI in the tabular case can be summarized as Algorithm \ref{alg:CPI:tabular}. 
\begin{algorithm}[!htb]
\caption{CPI Algorithm: tabular case}
\label{alg:CPI:tabular}
\begin{algorithmic}
\STATE Initial with arbitrary $\pi_0$, $V_0$
\STATE Initialize iteration index $K=0$
\STATE Given an arbitrarily small positive $\epsilon$
\REPEAT
\STATE Rollout $N$ steps from $\forall x_k\in\Omega$ with policy $\pi_K$\\
\STATE Receive and store $x_{k+i}$, $i\in[1,N]$
\STATE Policy evaluation: 

\setlength{\leftskip}{0.5em}
\REPEAT
\STATE Calculate $G(x_k,\pi_K,V_{K,q})$ using \eqref{eq.target_return}
\STATE Update value function using \begin{equation}   
\label{eq.bellmanupdate}
V_{K,q+1}(x_k) = G(x_k,\pi_K,V_{K,q}),\quad \forall x_k \in \Omega
\end{equation}
\UNTIL $\|V_{K,q+1}(x_k) - V_{K,q}(x_k)\|_{\infty} \le \epsilon$ 
\STATE $V_{K+1}(x_k)=V_{K,q+1}(x_k),\quad \forall x_k \in \Omega$

\setlength{\leftskip}{0em}
\STATE Constrained policy improvement:
\STATE \quad Update policy using:\begin{equation} 
\label{eq.appendixpolicyopti}
\begin{aligned}
\pi_{K+1}(x_k)&=\arg\min_{\pi(x_k)} G(x_k,\pi,V_{K+1}) \\
s.t. \quad & x_{k+i+1} = f(x_{k+i},\pi(x_{k+i})), \quad i\in[0,N-1]\\
&  J_{C_\tau}(x_{k+i+1}) \le b_{\tau},  \qquad \quad \tau\in[1,\tau_{\text{max}}] 
\end{aligned}
\end{equation}
\STATE $K=K+1$
\UNTIL Convergence  
\end{algorithmic}
\end{algorithm}

Next, the  convergence property of Algorithm \ref{alg:CPI:tabular} will be established. As the iteration index $K\rightarrow\infty$, we will show that the optimal value function and optimal policy can be achieved using Algorithm \ref{alg:CPI:tabular}. Before the main theorem, some lemmas are necessary at this point. Lemma \ref{lemma.policyevalution} shows the convergence of policy evaluation, which has been proved and described in previous studies.
\begin{lemma} 
\label{lemma.policyevalution}
(Policy Evaluation \cite{sutton2018reinforcement}). For a fixed policy $\pi_K$, consider the Bellman backup rule in \eqref{eq.bellmanupdate} and a mapping $V_{K,0}(x_k): \mathbb{R}^n \rightarrow \mathbb{R}$, then the sequence $V_{K,q}(x_k)$ will converge to the value function $V^{\pi_K}(x_k)$ for $\forall x_k \in \Omega$ as $q\rightarrow \infty$.
\end{lemma}
\begin{proof}
Note that throughout our computation, the value function $V(x_k)$ is always bounded for $\forall x_k \in \Omega$ since $\gamma \in (0,1)$ and $l(x_k,u_k)$ is bounded. Define a maximum norm on the value function as
\begin{equation}
\nonumber
 \|V_{K,q+1}(x_k) - V_{K,q}(x_k)\|_{\infty} \doteq \max_{x_k\in\Omega}|V_{K,q+1}(x_k) - V_{K,q}(x_k)|.
\end{equation}
Suppose $\Delta = \|V_{K,1}(x) - V_{K,0}(x)\|_{\infty}$. According to \eqref{eq.target_return} and \eqref{eq.bellmanupdate}, it follows that
\begin{equation}   
\nonumber
\begin{aligned}
|V_{K,2}(x_k)&-V_{K,1}(x_k)| 
\\&= |G(x_k,\pi_K,V_{K,1})-G(x_k,\pi_K,V_{K,0})|
\\&= \gamma^N|V_{K,1} (x_{k+N}) - V_{K,0} (x_{k+N})|
\\&\le \gamma^N \Delta, \quad \forall x_k \in \Omega.
\end{aligned}
\end{equation}
Extending this for all subsequent iteration steps $q$, one has
\begin{equation}   
\nonumber
\begin{aligned}
|V_{K,q+1}(x_k)&-V_{K,q}(x_k)|
\\&=\gamma^N|V_{K,q}(x_{k+N}) - V_{K,q-1} (x_{k+N})| \\
&\le \gamma^{qN} \Delta, \quad \forall x_k \in \Omega.
\end{aligned}
\end{equation}
Therefore, $\lim_{q\rightarrow\infty}\|V_{K,q+1}(x)-V_{K,q}(x)\|_{\infty} = 0$, which also means that $V_{K,\infty}(x_k)$ satisfies the Bellman equation \eqref{eq.bellman} for $\forall x_k \in \Omega$. So, the sequence $V_{K,q}(x_k)$ converges to the value function $V^{\pi_K}(x_k)$ for $\forall x_k \in \Omega$ as $q\rightarrow \infty$, i.e., $V_{K+1}=V^{\pi_{K}}$.
\end{proof}

The following lemma shows that, the new policy $\pi_{K+1}$ obtained by \eqref{eq.appendixpolicyopti} has a lower value function $V^{\pi_{K+1}}$ than the old policy $V^{\pi_K}$. The proof borrows heavily from the policy improvement theorem of Q-learning and soft Q-learning \cite{sutton2018reinforcement,Haarnoja2018SAC,Haarnoja2018ASAC,haarnoja2017soft-Q}.

\begin{definition}
\label{def.feasible}
(Feasible State and Policy). A state $x_k$ is defined as a feasible state, denoted by  $x_k \in \Psi$ with respect to system \eqref{eq.statefunction}, if there is a policy that can ensure the system satisfies all the state constraints described in \eqref{eq.appendixpolicyopti}. If a policy $\pi$ works for $\forall x_k \in \Psi$, it is defined as a feasible policy, denoted by $\pi \in \Pi$.
\end{definition}

\begin{lemma} 
\label{lemma.policyimprovement}
(Constrained Policy Improvement). Suppose $\Omega \subseteq \Psi$. Assume that if $x_k\in \Psi$ and $\pi \in \Pi$, then state $x_{k+1}=f(x_k,\pi(x_k)) \in \Psi$ with respect to system \eqref{eq.statefunction}. Given the associated value function $V^{\pi_{K}}(x_k)$ of a policy $\pi_K(x_k)$, and the new policy $\pi_{K+1}(x_k)$ is obtained by solving \eqref{eq.appendixpolicyopti} for $\forall x_k \in \Omega$. Then $V^{\pi_{K+1}}(x_k)\le V^{\pi_K}(x_k)$ for $\forall x_k \in \Omega$ and $\forall K \in \mathbb{N}_+$.
\end{lemma}

\begin{proof}
Because $\Omega \subseteq \Psi$, \eqref{eq.appendixpolicyopti} is feasible for $\forall x_k \in \Omega$. Then, from \eqref{eq.appendixpolicyopti}, one has 
\begin{equation}  
\nonumber
G(x_k,\pi_K,V_{K+1}) \ge G(x_k,\pi_{K+1},V_{K+1}),\ \forall x_k \in \Omega.
\end{equation}
By Lemma \ref{lemma.policyevalution}, one has $V_{K+1}=V^{\pi_K}$. Furthermore, from \eqref{eq.target_return}, it is clear that
\begin{equation}  
\nonumber
\begin{aligned}
G(x_k,\pi_K,V_{K+1})=V^{\pi_K}(x_k)
,\ \forall x_k \in \Omega.
\end{aligned}
\end{equation}
Therefore, we can show that
\begin{equation}   
\label{eq.appendixpi}
\begin{aligned}
&V^{\pi_K}(x_k)\\ &\quad=\sum_{i=0}^{N-1}\gamma^{i}l(x'_{k+i},\pi_K(x'_{k+i}))+\gamma^N V^{\pi_K}(x'_{k+N}) \\
&\quad\ge \sum_{i=0}^{N-1}\gamma^{i}l(x_{k+i},\pi_{K+1}(x_{k+i}))+\gamma^N V^{\pi_K}(x_{k+N}) \\
&\quad\ge \sum_{i=0}^{2N-1}\gamma^{i}l(x_{k+i},\pi_{K+1}(x_{k+i}))+\gamma^{2N} V^{\pi_K}(x_{k+2N}) \\
&\quad\vdots\\
&\quad\ge \sum_{i=0}^{\infty} \gamma^{i}l(x_{k+i},\pi_{K+1}(x_{k+i}))\\
&\quad=V^{\pi_{K+1}}(x_k),  \quad \forall x_k\in\Omega,
\end{aligned}
\end{equation}
where $x'_{k+i} \sim \pi_K$, $x_{k+i} \sim \pi_{K+1}$ and $x'_k=x_k$. We have thus proven that 
\begin{equation}   
\nonumber
V^{\pi_{K+1}}(x_k) \le V^{\pi_K}(x_k), \quad \forall x_k \in \Omega.
\end{equation}
\end{proof}

\begin{theorem}
\label{theorem.GPI:tabular}
(Constrained Policy Iteration in Tabular Case). Through Algorithm \ref{alg:CPI:tabular}, any policy $\pi_0$ will converge to the global optimal policy $\pi^* \in \Pi$, such that $V^{\pi^*}(x_k) \le V^{\pi}(x_k)$ for $\forall \pi \in \Pi$ and $\forall x_k \in \Omega$.
\end{theorem}

\begin{proof} 
For the policy $\pi_K$ at iteration $K$, we can find its associated $V^{\pi_K}$ through policy evaluation process follows from Lemma \ref{lemma.policyevalution}. By Lemma \ref{lemma.policyimprovement}, $V^{\pi_{K+1}}(x_k) \le V^{\pi_K}(x_k)$ for $\forall x_k \in \Omega$. Besides, since $V^{\pi_K}(x_k)$ is bounded below by $V^{*}(x_k)$, $\pi_K$ and $V^{\pi_K}$ will converge to some $\pi_{\infty}$ and $V^{\pi_{\infty}}$. At convergence, for $\forall \pi \in \Pi$ and $x_k \in \Omega$, it must follow that $G(x_k,\pi_{\infty},V^{\pi_{\infty}})\le G(x_k,\pi,V^{\pi_{\infty}})$. Using the same iterative argument as in \eqref{eq.appendixpi} of Lemma \ref{lemma.policyimprovement}, it is clear that $V^{\pi_{\infty}}(x_k) \le V^{\pi}(x_k)$ for $\forall \pi \in \Pi$ and $x_k \in \Omega$. From \eqref{eq.optimal_V}, one has $V^{\pi_{\infty}}(x_k) = V^{*}(x_k)$.
Hence $\pi_{\infty}$ is optimal in $\Pi$, i.e., $\pi_{\infty}=\pi^*$.
\end{proof}

\subsection{Convergence Proofs of Constrained Policy Iteration with Function Approximation}
\label{sec.appendix_CPI_function}
The pseudo-code of CPI with function approximation is described as Algorithm \ref{alg:CPI:function}. By Lemma \ref{lemma.approximation} and \ref{lemma.global_min}, the convergence results of tabular CPI (Algorithm \ref{alg:CPI:tabular}) can be extended to Algorithm \ref{alg:CPI:function}.  

\begin{algorithm}[!htb]
\caption{CPI Algorithm: function approximation}
\label{alg:CPI:function}
\begin{algorithmic}
\STATE Initial with arbitrary $\theta_0$, $\omega_0$, learning rates $\alpha_c$
\STATE Given an arbitrarily small positive $\epsilon$
\STATE Initialize iteration index $K=0$
\REPEAT
\STATE Rollout $N$ steps from $\forall x_k\in\Omega$ with policy $\pi_{\theta_K}$\\
\STATE Receive and store $x_{k+i}$, $i\in[1,N]$
\STATE Policy evaluation: 

\setlength{\leftskip}{1em}
\REPEAT
\STATE \quad$\omega_{K,q+1}=\omega_{K,q}$
\STATE \quad Calculate $G(x_k,\pi_{\theta_K},V_{\omega_{K,q+1}})$ using \eqref{eq.target_return}

\setlength{\leftskip}{1em}
\REPEAT
\STATE \quad Calculate $\frac{\text{d}L(\omega_{K,q+1})}{\text{d}\omega_{K,q+1}}$ using \eqref{eq.value_update} 
\STATE \quad Update value function using \begin{equation}
\label{eq.appenupdaterule_V}
\omega_{K,q+1} = -\alpha_c \frac{\text{d}L(\omega_{K,q+1})}{\text{d}\omega_{K,q+1}}+\omega_{K,q+1}
\end{equation}
\UNTIL $L(\omega_{K,q+1}) \le \epsilon$

\setlength{\leftskip}{1em}
\UNTIL $\|V(x_k;\omega_{K,q+1}) - V(x_k;\omega_{K,q})\|_{\infty} \le \epsilon$ 
\STATE $\omega_{K+1}=\omega_{K,q+1}$

\setlength{\leftskip}{0em}
\STATE Constrained policy improvement: 
\STATE \quad  Construct $J(\theta)$ under $V_{\omega_{K+1}}$ using \eqref{eq.policy_objective}
\STATE \quad Update policy using \eqref{eq.constrained_PI}
\STATE $K=K+1$
\UNTIL Convergence
\end{algorithmic}
\end{algorithm}

\begin{lemma} 
\label{lemma.approximation}
(Universal Approximation Theorem \cite{Hornik1990Universal}). For any continuous function $F(x):\mathbb{R}^n\rightarrow\mathbb{R}^d$ on a compact set $\Omega$, there exists a feed-forward NN, having only a single hidden layer, which uniformly approximates $F(x)$ and its gradient to within arbitrarily small error $\epsilon \in \mathbb{R}_{+}$ on $\Omega$.
\end{lemma}

\begin{lemma}
\label{lemma.global_min} (Global Minima of Over-Parameterized Neural Networks \cite{allen2018convergence,du2018gradient}).
Consider the following optimization problem
\begin{equation}
\nonumber
\min_{\psi}\mathcal{L}(\psi) = \mathop{\mathbb{E}}_{X_i\in\mathcal{B}}\Big\{\frac{1}{2}(\mathcal{F}(X_i;\psi)-Y_i)^2\Big\},
\end{equation}
where $X_i\in \mathbb{R}^n$ is the training input, $Y_i \in \mathbb{R}^d$ is the associated label, $\mathcal{B}=\{(X_1,Y_1),(X_2,Y_2),\hdots\}$ is the dataset, $\psi$ is the parameter to be optimized, and $\mathcal{F}:\mathbb{R}^n \rightarrow \mathbb{R}^d$ is an NN. If the NN $\mathcal{F}(X;\psi)$ is over-parameterized (i.e., the number of hidden neurons and layers is sufficiently large), simple algorithms such as Gradient Descent (GD) or SGD can find global minima on the training objective $\mathcal{L}(\psi)$ in polynomial time, as long as the dataset $\mathcal{B}$ is non-degenerate. The dataset is non-degenerate if the same inputs $X_1=X_2$ have the same labels $Y_1=Y_2$.
\end{lemma}

\begin{theorem}
\label{theorem.GPI_approximation} 
(Constrained Policy Iteration with Function Approximation). Suppose both $V(x_k;\omega)$ and $\pi(x_k;\theta)$ are over-parameterized. Through Algorithm \ref{alg:CPI:function}, any initial parameters $\omega_0$ and $\theta_0$ will converge to $\omega^*$ and $\theta^*$, such that $V(x_k;\omega^*)=V^{\pi_{\theta^*}}(x_k) \le V^{\pi_{\theta}}(x_k)$ for $\forall \pi_{\theta} \in \Pi$ and $\forall x_k \in \Omega$. 
\end{theorem}

\begin{proof}
In the policy evaluation step of Algorithm \ref{alg:CPI:function}, by Lemma \ref{lemma.approximation}, there always $\exists \omega_{K,q+1}\in\mathbb{R}^p$, such that
\begin{equation}
\label{eq.appenbellman}
V(x_k;\omega_{K,q+1})=G(x_k,\pi_{\theta},V_{\omega_{K,q}}), \quad \forall x_k \in \Omega.
\end{equation}
In other words, $\exists \omega_{K,q+1}\in\mathbb{R}^p$, such that $L(\omega_{K,q+1}) \le \epsilon$. Since Algorithm \ref{alg:CPI:function} updates $\omega_{K,q}$ using \eqref{eq.appenupdaterule_V} to continuously minimize $L(\omega_{K,q})$, according to Lemma \ref{lemma.global_min}, we can find $\omega_{K,q+1}$ in polynomial time. From Lemma \ref{lemma.policyevalution}, as \eqref{eq.appenbellman} and \eqref{eq.bellmanupdate} are equivalent, the sequence $V(x_k;\omega_{K,q+1})$ converges to the value function $V^{\pi_{\theta_K}}(x_k)$ for $\forall x_k \in \Omega$ as $q\rightarrow \infty$, i.e., 
\begin{equation}
\label{eq.appen_pe_cpi}
V(x_k;\omega_{K+1})=V(x_k;\omega_{K,\infty})=V^{\pi_{\theta_K}}(x_k), \ \forall x_k \in \Omega.    
\end{equation}

For the policy improvement process, according to Lemma  \ref{lemma.approximation} and extending Lemma \ref{lemma.policyimprovement} to the function approximation case, one has
\begin{equation}
\label{eq.appen_pim_cpi}
V^{\pi_{\theta_{K+1}}}(x_k) \le V^{\pi_{\theta_{K}}}(x_k), \quad \forall x \in \Omega.
\end{equation}

Finally, according to \eqref{eq.appen_pe_cpi}, \eqref{eq.appen_pim_cpi} and Theorem  \ref{theorem.GPI:tabular}, we can get the conclusion that  $V^{\pi_{\theta_{\infty}}}(x_k) = V(x_k;\omega_{\infty}) = V^{*}(x_k)$ for $\forall x_k \in \Omega$, i.e., $\theta_{\infty}=\theta^*$ and $\omega_{\infty}=\omega^*$.
\end{proof} 

\subsection{The Generalized Policy Iteration Framework}
\label{sec.appendix_CGPI}
Algorithms based on PI framework, such as Algorithm \ref{alg:CPI:tabular} and \ref{alg:CPI:function}, proceed by alternately updating the value and policy functions. Note that while one function is being updated, the other remains unchanged. Besides, taking the policy evaluation process of Algorithm \ref{alg:CPI:tabular} and \ref{alg:CPI:function} as an example, each function usually requires multiple updating iterations to meet the terminal conditions, which is the so-called protracted iterative computation problem \cite{sutton2018reinforcement}. This often leads to slow learning. Therefore, for practical applications, almost all ADP or RL algorithms build on the GPI framework, which truncates the policy evaluation and policy improvement processes into an arbitrary step or even one-step update \cite{liu2017ADP,sutton2018reinforcement}. In recent years,  many experimental results and theoretical proofs have shown that almost all ADP or RL algorithms based on PI can be adapted to a GPI version without losing the convergence guarantees \cite{duan2019generalized,mnih2015human,duan2020DSAC,lillicrap2015DDPG,mnih2016asynchronous,Haarnoja2018SAC,Haarnoja2018ASAC,schulman2017PPo,haarnoja2017soft-Q,barth-maron2018D4PG}. Based on this fact and Theorem \ref{theorem.GPI_approximation}, we make the following remark. 
\begin{remark}
\label{remark.CGPI} 
(Constrained Generalized Policy Iteration). Suppose both $V(x_k;\omega)$ and $\pi(x_k;\theta)$ are over-parameterized. Thr-ough Algorithm \ref{alg:CGPI}, any initial parameters $\omega_0$ and $\theta_0$ will converge to $\omega^*$ and $\theta^*$, such that $V(x_k;\omega^*)=V^{\pi_{\theta^*}}(x_k) \le V^{\pi_{\theta}}(x_k)$ for $\forall \pi_{\theta} \in \Pi$ and $\forall x_k \in \Omega$.
\end{remark}

\section{Acknowledgment}
This study is supported by the Beijing Science and Technology Plan Project with Z191100007419008, Tsinghua University-Didi Joint Research Center for Future Mobility, and NSF China with 51575293 and U20A20334. We would like to acknowledge Jie Li, Hao Chen, Yang Zheng, Yiwen Liao, Guofa Li, Ziyu Lin and Jiatong Xu for their valuable suggestions. The authors are grateful to the Editor-in-Chief, the Associate Editor, and anonymous reviewers for their valuable comments.

\printcredits

\bibliographystyle{model1-num-names}

\bibliography{cas-refs}

\end{document}